\documentclass[fleqn,usenatbib]{mnras}

\usepackage{newtxtext,newtxmath}

\usepackage[T1]{fontenc}

\DeclareRobustCommand{\VAN}[3]{#2}
\let\VANthebibliography\thebibliography
\def\thebibliography{\DeclareRobustCommand{\VAN}[3]{##3}\VANthebibliography}

\usepackage{graphicx}	
\usepackage{amsmath}	

\newcommand{\bs}[1]{\boldsymbol{#1}}\defcitealias{Sormani2022}{S22}
\usepackage{xcolor}
\definecolor{pink}{rgb}{0.96, 0.76, 0.76}

\usepackage{paralist}

\title[Bar formation epoch from NSD Mira variables]{The epoch of the Milky Way's bar formation: dynamical modelling of Mira variables in the nuclear stellar disc}

\author[J. L. Sanders et al.]{Jason L. Sanders,$^{1}$\thanks{E-mail: jason.sanders@ucl.ac.uk (JLS)}
Daisuke Kawata,$^{2}$
Noriyuki Matsunaga,$^{3}$
Mattia C. Sormani,$^{4}$
Leigh C. Smith,$^{5}$\newauthor
Dante Minniti,$^{6,7,8}$
Ortwin Gerhard$^{9}$
\\
$^{1}$Department of Physics and Astronomy, University College London, London WC1E 6BT, UK\\
$^{2}$Mullard Space Science Laboratory, University College London, Holmbury St. Mary, Dorking, Surrey RH5 6NT, UK\\
$^{3}$Department of Astronomy, School of Science, The University of Tokyo, 7-3-1, Hongo, Bunkyo-ku, Tokyo 113-0033, Japan\\
$^{4}$School of Mathematics $\&$ Physics, University of Surrey, Guildford, GU2 7XH, UK\\
$^{5}$Institute of Astronomy, University of Cambridge, Madingley Rise, Cambridge, CB3 0HA, UK\\
$^{6}$Departamento de Ciencias F\'isicas, Facultad de Ciencias Exactas, Universidad Andr\'es Bello, Fern\'andez Concha 700, Las Condes, Santiago, Chile\\
$^{7}$Vatican Observatory, Vatican City State, V-00120, Italy\\
$^{8}$Dept. of Physics, Universidade Federal de Santa Catarina, Florianopolis, Brazil\\
$^{9}$Max-Planck Institut f\"ur extraterrestrische Physik, Giessenbachstra\ss e, 85748 Garching, Germany
}

\date{Accepted XXX. Received YYY; in original form ZZZ}

\pubyear{2023}

\begin{document}
\label{firstpage}
\pagerange{\pageref{firstpage}--\pageref{lastpage}}
\maketitle

\begin{abstract}
A key event in the history of the Milky Way is the formation of the bar. This event affects the subsequent structural and dynamical evolution of the entire Galaxy. When the bar formed, gas was likely rapidly funnelled to the centre of the Galaxy settling in a star-forming nuclear disc. The Milky Way bar formation can then be dated by considering the oldest stars in the formed nuclear stellar disc. In this highly obscured and crowded region, reliable age tracers are limited, but bright, high-amplitude Mira variables make useful age indicators as they follow a period--age relation.
We fit dynamical models to the proper motions of a sample of Mira variables in the Milky Way's nuclear stellar disc region. Weak evidence for inside-out growth and both radial and vertical dynamical heating with time of the nuclear stellar disc is presented suggesting the nuclear stellar disc is dynamically well-mixed. Furthermore, for Mira variables around a $\sim350$ day period, there is a clear transition from nuclear stellar disc-dominated kinematics to background bar-bulge-dominated kinematics.
Using a Mira variable period--age relation calibrated in the solar neighbourhood, this suggests the nuclear stellar disc formed in a significant burst in star formation $(8\pm 1)\,\mathrm{Gyr}$ ago, although the data are also weakly consistent with a more gradual formation of the nuclear stellar disc at even earlier epochs.
This implies a relatively early formation time for the Milky Way bar ($\gtrsim8\,\mathrm{Gyr}$), which has implications for the growth and state of the young Milky Way and its subsequent history.
\end{abstract}



\section{Introduction}

With the advance of large astrometric, photometric and spectroscopic stellar surveys, a detailed picture and understanding of the formation and evolution of the Milky Way is being built up. About $13\,\mathrm{Gyr}$ ago, the proto-disc of the Milky Way likely formed from an early turbulent, merger-dominated phase in which it was rapidly metal-enriched and spun up \citep{Arentsen2020, BelokurovKravtsov2022, Rix2022}. Since then the Milky Way has undergone a series of merger events: about $10\,\mathrm{Gyr}$ ago \citep{Belokurov2020, Bonaca2020}, the Gaia-Sausage-Enceladus galaxy \citep{Belokurov2018, Helmi2018} merged with the Milky Way, and today the Sagittarius dwarf galaxy and Magellanic system are merging with the Milky Way in processes that have likely produced the non-equilibrium structure and features we see throughout the disc and the halo \citep{Antoja2018, Petersen2021, Drimmel2023, Koposov2023}.

A further significant event must be placed within this chronology of the Galaxy: the formation epoch of the Galactic bar. It is generally accepted that the Milky Way has a bar-bulge that evolved from a pre-existing disc \citep{BlitzSpergel1991, WeggGerhard2013}. This likely occurred in two distinct stages: a flattish in-plane bar was initially produced through an in-plane disc instability \citep{Hohl1971, Ostriker1973, SparkeSellwood1987} that may have been triggered by an external perturbation from a satellite galaxy \citep{Noguchi1987, Gerin1990, Lokas2014, Lokas2021}, and then the central parts fattened into the observed bar-bulge either through a violent buckling event \citep{Raha1991}, or a more gradual resonant thickening process \citep{Combes1981, Combes1990, Quillen2014, SellwoodGerhard2020}. 

Since the epoch of bar formation, the Milky Way bar has likely had a significant dynamical and evolutionary impact on the Galaxy in several ways:
\begin{inparaenum}
    \item The barred potential has likely led to significant restructuring of the Galactic disc. This is most notable through the presence of moving groups in the solar neighbourhood that have been related to various resonances in the bar \citep{Dehnen2000, Antoja2014, Monari2017, PerezVillegas2017, Hunt2018, Monari2019, Chiba2021a}, but it is also likely that the resonant overlap between the bar and spirals has led to enhanced radial migration \citep{Quillen2003, Minchev2010} that has shuffled stars radially in the disc \citep{Frankel2020};
    \item The dark matter halo has likely exchanged angular momentum with the Galactic bar through a dynamical friction process causing the bar to slow  \citep{TremaineWeinberg1984, Athanassoula2003}. This is an excellent probe of the fundamental properties of dark matter and is suggested by the morphology and metallicity distribution of the Hercules moving group in the solar neighbourhood \citep{Chiba2021b}. However, it is also known that both gas and tidal interactions can cause the bar to spin up again \citep[e.g.][]{Lokas2014, Beane2022};
    \item Finally, the presence of a bar significantly alters the structure of gas orbits within a Galaxy \citep{Binney1991, Fux1999, Sormani2015a, Sormani2015b, Li2022}, and in particular can cause gas to be directly funnelled towards the central regions of the Galaxy \citep{BournaudCombes2002, Hatchfield2021, SormaniBarnes}. It is therefore likely that bar formation is somehow linked to active galactic nucleus activity and black hole growth within the Milky Way but this is currently unclear \citep{Shlosman1989, SellwoodReview, Emsellem2015}.
\end{inparaenum} 
In addition to these evolutionary effects, bars only form in galaxies that are sufficiently kinematically cold and baryon-dominated \citep{Ostriker1973, Hohl1976, Athanassoula2002, Fragkoudi2021, BlandHawthorn2023}. In this way, a measurement of the bar age gives a direct probe of the morphological properties of the Milky Way and its relative dark matter/baryon fraction at a fixed epoch, allowing for detailed testing of the cosmological picture \citep[e.g.][]{Sheth2012}.

\subsection{When do bars form in galaxies?}
To understand these phenomena and effects further, it is crucial to date the formation of the Milky Way bar, making it a key goal of Galactic archaeology. However, this task is not simple, as, importantly, the dynamical age of a galaxy's bar can be different from the age of the constituent stars: stars can be born in a precursor disc from which the bar forms, stars can form in the bar region or a bar can capture stars that formed after bar formation as it grows and evolves. From observing galaxies across redshift, constraints can be placed on the occurrence of bars in galaxies and the maximum redshift at which galaxies host bars. This information gives the typical age of bars in galaxies other than the Milky Way. Locally, approximately two-thirds of spiral galaxies are observed to host bars \citep{Eskridge2000, Erwin2018}. Studies using Hubble Space Telescope (HST) data \citep{Sheth2008, Melvin2014, Simmons2014} have found that the fraction of \emph{strongly} barred galaxies decreases with redshift. However, the rate of this decrease may in part arise from the detectability of bars at high redshift with HST as recent JWST analyses \citep{Chen2022, Ferreira2022, Guo2023, Jacobs2023} have found several examples of barred galaxies at look-back times between $8-11\,\mathrm{Gyr}$ and a recent population-level analysis \citep{LeConte2023} finds a factor $3-4$ higher bar fractions at $z\gtrsim 1$ than the earlier HST results. Such observations suggest that the conditions are right for bars to form quite early on in the typical history of galaxies \citep{BlandHawthorn2023}, as indicated by the analysis of the bar fractions in cosmological simulations \citep{Fragkoudi2020, RosasGuevara2023}. However, it should be noted that the results on bars from cosmological simulations are likely still in a state of flux as there are tensions with the data in the distribution of bar lengths and/or pattern speeds they predict \citep{Zhao2020, Roshan2021, Fragkoudi2021, Frankel2022}.

There are several estimates of the epoch of bar formation in the Milky Way. The Galactic bar contains predominantly old stars of around $\sim10\,\mathrm{Gyr}$ \citep{Bernard2018, Bovy2019, Surot2019, Savino2020, Hasselquist2020, Grady2020} approximately supporting the idea that the Milky Way bar is old. However, there is the suggestion that the bar is not solely composed of old stars: using microlensed dwarf stars towards the Galactic bar-bulge, \cite{Bensby2013, Bensby2017} argued that there is $\sim20$ per cent of stars younger than $\sim5\,\mathrm{Gyr}$ old, which is approximately corroborated by the tail towards younger ages observed in HST colour-magnitude diagram modelling of the Galactic bar-bulge \citep{Bernard2018}. This minority of stars may be linked with in-bar star formation \citep[e.g.][]{Anderson2020} or capture. Based on the occurrence of photometrically-identified carbon stars in the Galactic bar-bulge using 2MASS, \cite{ColeWeinberg2002} argued the bar might be as young as $3-6\,\mathrm{Gyr}$ old, although \cite{Matsunaga2017} suggested that the photometric classification \citeauthor{ColeWeinberg2002} employed introduced highly-reddened older oxygen-rich stars. It seems that the low number of genuine carbon-rich AGB stars in the bar-bulge is consistent with their production through binary evolution \citep{SandersMatsunaga2023} which would push the bar age estimate from \cite{ColeWeinberg2002} higher. Recently, \cite{TepperGarcia2021} presented a simulation of a Milky Way-like galaxy that they argued matches the Milky Way well in several aspects and then supports the idea that the bar formed $3-4\,\mathrm{Gyr}$ ago, but similarly \cite{Buck2018} presented a Milky Way analogue simulation with a bar formation time of $\sim8\,\mathrm{Gyr}$. Using the potential dynamical impact of the bar, \cite{Khoperskov2019} presented a scenario for the formation of the `Gaia snail', the local spiral in the vertical displacement vs. vertical velocity of the stars \citep{Antoja2018}, in which the perturbation arises from the buckling of the bar. Although there are alternatives to their presented scenario (such as a satellite perturbation), this suggests a bar buckling event within the last $\sim 3\,\mathrm{Gyr}$, and if bars buckle promptly, possibly then a more recent bar formation. Finally, \cite{Wylie2022} used APOGEE data to identify an inner $\sim4\,\mathrm{kpc}$ ring of metal-rich stars containing red giants of around $4-9\,\mathrm{Gyr}$ old with a peak at $7\,\mathrm{Gyr}$, thus putting a lower limit on the bar formation epoch of $7\,\mathrm{Gyr}$ ago.

\subsection{Bar formation and nuclear stellar discs}
Another very different approach to dating the epoch of bar formation is to not look at the stars in the bar at all but instead to look at stars in structures that probably only started forming when bars have formed: nuclear stellar discs \citep[NSDs,][]{Erwin2002, Pizzella2002}. Controlled simulations of disc galaxies \citep[e.g.][]{Athanassoula1992, Cole2014, Seo2019, Baba2019} have demonstrated that when a bar forms, gas funnels along bar lanes towards the centre of the galaxy where it can settle in a nuclear ring (Sormani, Sobacchi and Sanders, in prep.) and begin forming stars that constitute a nuclear disc. Based on these theoretical ideas, the MUSE-TIMER project \citep{Gadotti2015, Gadotti2018, Gadotti2020,deSaFreitas2023,deSaFreitas2023b} has measured the age and abundance distributions of the nuclear regions of barred galaxies, finding that nuclear discs typically have lower velocity dispersions, are more metal-rich and are younger than the surrounding bar, with the indication of inside-out growth of the NSDs \citep{Bittner2020}.

This approach to measuring the formation age of a bar was highlighted specifically for dating the Milky Way bar formation epoch by \cite{Baba2019}. They demonstrated from controlled simulations that within $\sim1\,\mathrm{Gyr}$ of a bar forming there is an associated $\sim1\,\mathrm{Gyr}$ long central star formation burst forming an NSD and highlighted the importance of proper motion data for isolating the NSD population. The Milky Way is known to host an NSD of around $10^9\, M_\odot$ with a scalelength of $\sim75\,\mathrm{pc}$ and a scaleheight of $\sim25\,\mathrm{pc}$ that rotates at approximately $100\,\mathrm{km\,s}^{-1}$ \citep{Catchpole1990, Lindqvist1992, Launhardt2002, Nishiyama2013, Schoenrich2015, Matsunaga2015, GallegoCano2020, Shahzamanian2021, Schultheis2021, Sormani2022}. The nuclear star cluster \citep[NSC,][]{Neumayer2020} sits at the centre of the NSD and has an effective radius of $\sim 5\,\mathrm{pc}$. Recently, \cite{Sormani2022} fitted self-consistent axisymmetric dynamical models to a combination of spectroscopic and proper motion data \citep{Fritz2021, VIRAC} for $\sim3500$ giant stars across the NSD. The assumption of axisymmetry gives very good fits to the data, although detecting whether the NSD is genuinely an axisymmetric disc or a nuclear bar is challenging \citep{Alard2001, Gonzalez2012, Gerhard2012}. There have been several studies of the Milky Way NSD star formation history, beginning with \cite{Figer2004} who found a relatively continuous star formation history using Hubble Space Telescope. More recently, the GALACTICNUCLEUS survey \citep{GALACTICNUCLEUS} has provided deep $JHK_s$ photometry from which the morphology of the NSD giant branch luminosity function can be analysed. These data present a richer picture suggesting that the star formation history of the central NSD is more bursty, predominantly forming $\gtrsim7-8\,\mathrm{Gyr}$ ago \citep[consistent with the presence of RR Lyrae stars in this region,][]{Minniti2016} with a more recent star formation burst in the last $\sim1\,\mathrm{Gyr}$ \citep{Matsunaga2015, Nogueras-Lara2020, Schoedel2023}. However, the outer parts of the NSD have evidence of more significant intermediate-age populations of $2-7\,\mathrm{Gyr}$ \citep{NoguerasLara2022, NoguerasLara2023}, which supports an inside-out growth of the NSD in the Milky Way.

\subsection{Mira variables as age tracers of the nuclear stellar disc}
One useful \emph{resolved} age tracer for the NSD region is Mira variable stars. These are high-amplitude (typically $\Delta K_s\gtrsim0.4$) asymptotic giant branch stars with periods of $80$ to $1000$ days thermally pulsing in the fundamental mode \citep{Wood2015} driven through some convective mechanism \citep{Freytag2017, Xiong2018}. Primarily from observations of the LMC, they are known to follow period--luminosity relations \citep{GlassLloydEvans1981, Feast1989, Ita2004, Groenewegen2004, Fraser2008, Riebel2010, Ita2011, Yuan2017, Yuan2018, Bhardwaj2019, Iwanek2021b, Sanders2023} as stars of a given mass only begin pulsating in the fundamental mode at a narrow range of radii \citep{Trabucchi2019}. For a similar reason, there is a theoretical expectation that Mira variables follow a period--age relation \citep{WyattCahn1983, FeastWhitelock1987, Eggen1998, Trabucchi2022}. Such a relation has been known observationally for some time from the related correlations between the periods of the Mira variables and their kinematics \citep{Merrill1923} and between the periods and the scale-heights of the populations \citep{Feast1963}. As hotter kinematics are typically associated with older populations, these correlations are typically interpreted as manifestations of a period--age relation, which is further observationally validated through the limited number of cluster Mira variables \citep{Grady2019, Marigo2022, ZhangSanders2023}. The solar neighbourhood correlations led to a number of empirical calibrations \citep{FeastWhitelock1987, FeastWhitelock2000P, Feast2006, FeastWhitelock2014, Catchpole2016, LopezCorredoira2017, Grady2020, Nikzat2022}. Recently, \cite{ZhangSanders2023} fitted dynamical models to Mira variables from Gaia DR3 and by relating the radial and vertical dispersions with period to the dispersions as a function of age from the analysis of main-sequence turn-off and sub-giant stars from \cite{Yu2018}, they derived a period--age relation that agreed well with both previous analyses and the results from the cluster members.

The calibrated period--age relations demonstrate that Mira variables span ages from $\sim1-10\,\mathrm{Gyr}$ making them excellent probes of intermediate age populations and so they have found significant use as an archaeological probe of the Milky Way \citep{Catchpole2016, Grady2019, Grady2020, Semczuk2022, ZhangSanders2023}. In addition to this, the brightness and low contamination of Mira variables also make them ideal age tracers of the NSD. Recently, \cite{Sanders2022} presented a sample of $1782$ Mira variables in the NSD region extracted from the VVV survey. This catalogue built on previous work searching for Mira variables by targeting OH/IR maser stars \citep[][which are biased towards longer period sources]{Blommaert1998, Wood1998} and the broader variable star searches of \cite{Glass2001} and \cite{Matsunaga2009} that only targeted smaller on-sky areas close to the Galactic Centre. In addition to providing a more panoramic view of the entire NSD region (the inner $3\times3\,\mathrm{deg}^2$), the catalogue of \cite{Sanders2022} contained proper motions for all Mira variable candidates from the VIRAC reduction of VVV \citep{VIRAC}. Although NSD membership can be assessed using Mira variable period--luminosity relations, these rely on a solid understanding of the extinction (both interstellar and circumstellar) and population effects, making kinematic membership significantly more reliable. With this sample, we are in a position to kinematically explore and identify the onset of star formation in the NSD and hence the epoch of bar formation in the Milky Way.

In this work, we present dynamical models of the Mira variable population in the NSD region from \cite{Sanders2022}. As contamination from the foreground disc and bar is significant \citep{Sormani2022}, the models consist of two components: an axisymmetric model for the NSD population and a barred `background' $N$-body model from \cite{Portail2017}. The NSD models and their relative weight compared to the background are parametrised by the period/age of the Mira variable population. Furthermore, we provide a detailed discussion and implementation of the selection function of the sample, which preferentially biases us towards observing fainter background objects. The paper is structured as follows: in Section~\ref{sec:data} we describe the dataset used, whilst in Section~\ref{sec:models} we lay out the components of the model providing a detailed discussion of the selection function of the sample in Appendix~\ref{appendix:selection_function}. The results of our modelling are presented in Section~\ref{sec:results} and our conclusions and a discussion of the implications of our results are described in Section~\ref{sec:discussion}.

\section{Data and initial modelling considerations}\label{sec:data}
\begin{figure}
    \centering
    \includegraphics[width=\columnwidth]{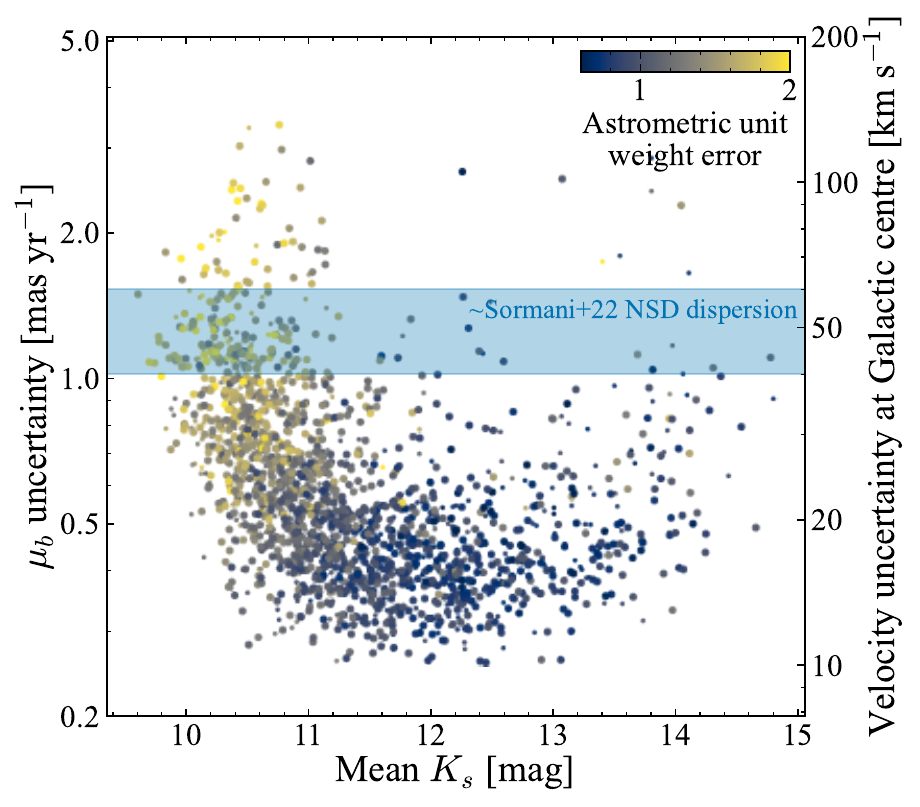}
    \caption{Astrometric data quality: Galactic latitudinal proper motion uncertainty ($\mu_b$) against VIRAC $K_s$ averaged over observed epochs coloured by astrometric unit weight error. The smaller points are unreliable or low amplitude. The band is the approximate vertical velocity dispersion of the NSD from \protect\cite{Sormani2022}.}
    \label{fig:uncert}
\end{figure}

The Mira variable sample is taken from \cite{Sanders2022}. This sample was extracted from an intermediate version of the VIRAC2 reduction of the VVV data \citep{VIRAC}.
VIRAC2 is the second iteration of the VVV \citep[Vista Variables in Via Lactea,][]{Minniti2010, Saito2012} InfraRed Astrometric Catalogue constructed from fitting five-parameter astrometric solutions to point-source-function catalogues of the VVV and VVVX $K_s$ epoch data over a baseline of around $10$ years that was photometrically calibrated to 2MASS data and astrometrically calibrated to the Gaia EDR3 reference frame \citep[e.g.][]{Sanders2019a, Clarke2019}. In this way, absolute proper motions are available despite the lack of background astrometric calibrators in this highly crowded and extincted region. Likely Mira variables were first identified in the inner $3\times3\,\mathrm{deg}^2$ of the Galaxy using basic cuts on the $K_s$ amplitude inspired by the study of \cite{Matsunaga2009}, and then more detailed light curve modelling of the selected stars was performed to measure periods allowing selection using the period--amplitude diagram and some period--Wesenheit index planes. Any Mira variable from \cite{Matsunaga2009} missed by the pipeline but still in the VIRAC catalogue is also included. There are $1782$ Mira variables in the sample, all of which have proper motion measurements. The $.5$th and $99.5$ percentile of period distribution are $105$ and $845$ days respectively. There are $21$ stars within the NSC region of $<r_\mathrm{eff}\approx5\,\mathrm{pc}$ from the Galactic Centre and $51$ within $2r_\mathrm{eff}$. The median proper motion uncertainty in each component is $\sim0.5\,\mathrm{mas\,yr}^{-1}$ corresponding to $\sim20\,\mathrm{km\,s}^{-1}$ at the Galactic centre distance of $8.275\,\mathrm{kpc}$ \citep{GravityCollaboration}. The full distribution of the uncertainties is shown in Fig.~\ref{fig:uncert}, which demonstrates the deterioration of the astrometry for the brighter stars due to saturation effects. The unit weight error calculated from the residuals to the astrometric fit increases from $<1$ for stars with $K_s\gtrsim11.5$ to $\sim 2$ for stars with $K_s\sim10.5$. This suggests that the astrometric centroid errors are underestimated for stars in the saturated regime and possibly the resulting proper motion uncertainties are underestimated. In addition to this, \cite{Luna2023} compared the VIRAC2 and HST proper motions in a few available fields finding that possibly VIRAC2 proper motion uncertainties are underestimated by $\sim10$ per cent. 
From the model fits of \cite{Sormani2022} the expected vertical velocity dispersion of the NSD is around $50\,\mathrm{km\,s}^{-1}$ (although note it decreases with radius and the velocity distributions are cuspier than a Gaussian such that the standard deviation does not capture the full distribution) which the proper motion uncertainties of our sample are small enough to resolve. $91$ Mira variables have been flagged as unreliable from a visual inspection of the light curves \citep{Sanders2022}. Furthermore, $342$ stars have $K_s$ amplitudes $<0.4$ making their Mira variable classification more suspect \citep{Matsunaga2009}. For some of the Mira variable candidates, radial velocities are available from maser observations (\citealt{Engels2015} for OH masers and \citealt{Messineo2002,Messineo2004}, \citealt{Deguchi2004} and \citealt{Fujii2006} for SiO masers).

\begin{figure*}
    \centering
    \includegraphics[width=\textwidth]{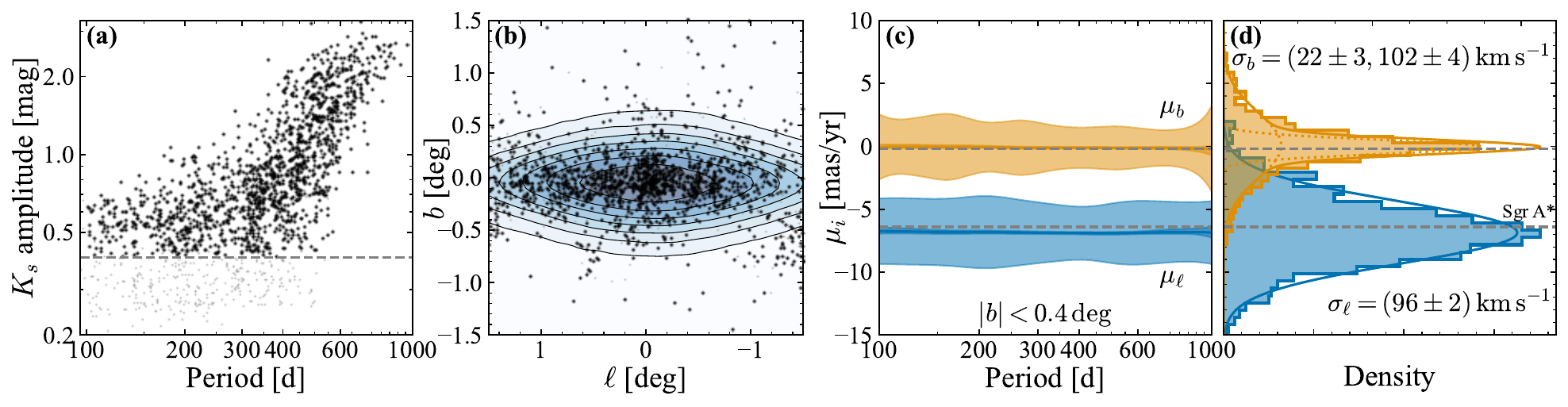}
    \caption{Summary of NSD Mira variable sample: panel (a) shows the period--amplitude distribution (unreliable and low-amplitude Mira variable candidates are shown with smaller grey points); panel (b) shows the distribution of reliable, high-amplitude Mira variable candidates on the sky with the background contours showing the ratio of the NSD surface density to the total (NSD + bar-bulge) surface density (excluding the NSC) from the models of \protect\cite{Sormani2022} (the contours are equally spaced every ten per cent with the outermost contour corresponding to ten per cent); panel (c) shows the mean proper motion, its uncertainty (darker bracket) and the standard deviation (lighter bracket) for the reliable, high-amplitude Mira variables with $|b|<0.4\,\mathrm{deg}$; panel (d) shows histograms and two-component Gaussian mixture models (solid lines are the full models and dotted lines each component) for the proper motions of the same sample. The proper motion of Sgr A* is marked by grey dashed lines in panels (c) and (d).}
    \label{fig:data}
\end{figure*}

We display the data sample in Fig.~\ref{fig:data}. The concentration of Mira variables towards the plane is a combination of the NSD flattening but perhaps more predominantly selection effects (see next section). We have fitted two simple models to the Galactic longitudinal and latitudinal proper motion distributions of the reliable high-amplitude sample with $|b|<0.4\,\mathrm{deg}$ ($58\,\mathrm{pc}$ projected height): (i) a Gaussian model with a flexible smoothed cubic spline as a function of period \citep[][with $10$ equally-spaced knots in $\log_{10}P$ between the $.5$th and $99.5$th percentile of the data]{jaxcosmo} for the mean and standard deviation, and (ii) a two-component Gaussian mixture model. Both models account for the formal uncertainties in the proper motions. The models are implemented in \textsc{Jax} \citep{jax} and \textsc{Numpyro} \citep{numpyro1,numpyro2}, and sampled using the NUTS sampler \citep{nuts_sampler}\footnote{See \url{https://adrian.pw/blog/flexible-density-model-jax/} for a useful description of using \textsc{Jax} and \textsc{Numpyro} with splines for stream modelling.}. In addition to removing the high-amplitude and unreliable stars, we opt to further remove $23$ stars with proper motions more than $3$ times the standard deviation away from the median so in total $1382$ stars are used. The results are shown in the right panels of Fig.~\ref{fig:data}. The mean proper motions are in accord with the motion of Sgr A*. The standard deviations are relatively constant with period but a slight narrowing is visible towards longer periods (particularly for $\mu_b$; note the broadening at long period occurs beyond the $99.5$th percentile of the sample). As assessed by the Bayesian information criterion, two components are not necessary for the $\mu_\ell$ distribution which is well represented by a single Gaussian with $\sim100\,\mathrm{km\,s}^{-1}$ dispersion \citep[assuming all stars are situated at the Galactic centre; note][suggest $\mu_\ell$ is well modelled by three Gaussian components]{Shahzamanian2021}. However, the $\mu_b$ distribution consists of two Gaussian components (with approximately equal weights): a colder core with a dispersion of $(22 \pm 3)\,\mathrm{km\,s}^{-1}$ and a hotter component with $(102 \pm 4)\,\mathrm{km\,s}^{-1}$. Using giant stars in a central field of $|\ell|\lesssim0.25\,\mathrm{deg}$ and $|b|\lesssim0.1\,\mathrm{deg}$, \citep{Shahzamanian2021} also find two Gaussian fits are appropriate with the hotter component approximately consistent with our findings. However, the colder component from \cite{Shahzamanian2021} has dispersion $\sim60\,\mathrm{km\,s}^{-1}$. This is broader than our findings here possibly due to the decay of the NSD vertical dispersion with radius, the non-Gaussianity of the NSD velocity distribution meaning the balance of bar and NSD with location impacts the results or the impact of background disc stars on our sample \citep{Sormani2022}. The mean velocity of the colder component is $(-2.0\pm1.7)\,\mathrm{km\,s}^{-1}$ in relatively good agreement with the expected reflex motion from the Sun's vertical velocity of $7.25\,\mathrm{km\,s}^{-1}$ \citep{Schoenrich2010} giving good evidence that the colder component is part of the NSD. 

\begin{figure}
    \centering
    \includegraphics[width=\columnwidth]{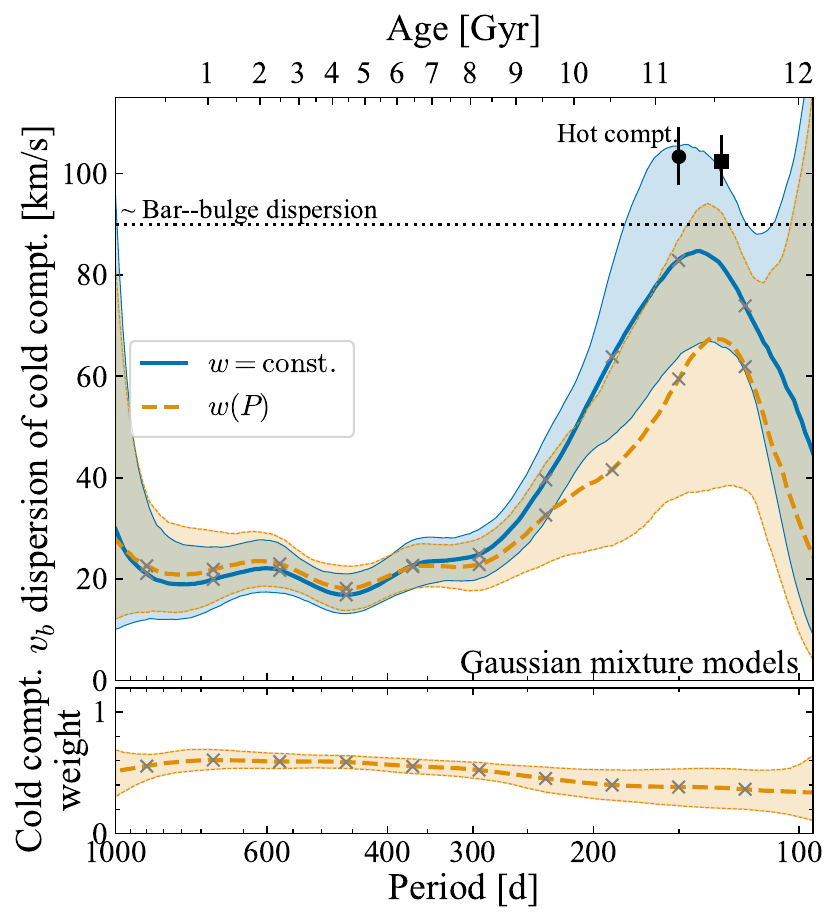}
    \caption{Flexible Gaussian mixture model fitting results to the latitudinal proper motion distributions, $\mu_b$. We display the dispersion of the colder component (converted into a vertical velocity assuming all stars are at the Galactic Centre distance of $8.275\,\mathrm{kpc}$) with crosses showing the spline knots. The blue solid line is for a model with a fixed cold component weight with period and the orange dashed is for a period-dependent cold component weight (as displayed in the lower panel). The brackets are $\pm1$ standard deviation. The hot component has dispersion given by the errorbars (circle for the fixed weight model and square for the variable; positioned at an arbitrary $x$-axis location). The approximate dispersion of the bar-bulge is given by the dotted line.}
    \label{fig:mix}
\end{figure}

We investigate further the narrowing of the dispersion with period by fitting Gaussian mixture models as a function of period. We fit two-component Gaussian mixture models where the standard deviation of the first Gaussian is a cubic spline with period whilst the second has a period-independent dispersion. We set the Gaussian means at the reflex solar velocity. The results are shown in Fig.~\ref{fig:mix} for two variants: one for fixed mixture weight with period and one with a cubic spline variation of the mixture weight with period (all splines are set to have $10$ knots logarithmically spaced between $120$ and $900$ days). We note how in both cases the dispersion is around $20\,\mathrm{km\,s}^{-1}$ at long period and then there is a transition around $300$ day period below which the dispersion rises towards that of the second component and the expected dispersion of the bar-bulge \citep[around $90\,\mathrm{km\,s}^{-1}$,][]{Sanders2019a}. 

The drawbacks of this simple approach are that (i) it does not capture spatial trends in kinematics, (ii) the velocity distributions are not Gaussian \citep{Sormani2022} and (iii) no selection function or line-of-sight distance distribution is considered and we effectively consider all of the stars at the Galactic centre distance. However, these initial considerations have clearly demonstrated that (i) some part of the data suggest NSD kinematics, (ii) there is a transition around a period of $\sim300$ day with longer period stars showing colder NSD kinematics and shorter period stars exhibiting more bar-bulge-like kinematics, and (iii) the contamination from the surrounding bar-bulge is significant, we will now turn to a more sophisticated modelling approach to address the highlighted drawbacks and elucidate some of these features of the data further.

\section{Multi-component dynamical modelling framework}\label{sec:models}
The aim is to model the transverse kinematics (proper motions, $\bs{\mu} = \bs{v}_\mathrm{t}/s$ for transverse velocity $\bs{v}_\mathrm{t}$ and distance $s$) of the Mira variable stars as a function of their on-sky locations (Galactic coordinates: $\ell,b$) and their periods, $P$. We are therefore building a likelihood from $p(\bs{\mu}|\ell,b, P)$. This is similar in spirit to \cite{ZhangSanders2023}. We begin by writing
\begin{equation}
    p(\bs{\mu}|\ell,b,P) = \int\mathrm{d}^2\tilde{\bs{\mu}}\,\mathrm{d}s\,\mathcal{N}(\bs{\mu}|\tilde{\bs{\mu}},\bs{\Sigma_\mu})S(\ell,b,s,P)p(\tilde{\bs{\mu}},s|\ell,b,P),
\label{eqn::highlevel}
\end{equation}
where $\mathcal{N}(\bs{\mu}|\tilde{\bs{\mu}},\bs{\Sigma_\mu})$ is a multivariate Gaussian accounting for proper motion uncertainties and $\tilde{\bs{\mu}}$ the true proper motions.
We neglect uncertainties in the period -- typically the posterior distribution from Lomb-Scargle/Fourier fitting methods is narrow or significantly multi-modal due to a series of alias peaks \citep[see figure C1 of][for tests of the period recovery quality of the sample]{Sanders2022}. $S(\ell,b,s, P)$ is the selection function which gives the fraction of stars at each Galactic coordinate $(\ell,b)$, distance $s$ and period $P$ that enters the sample. Appendix~\ref{appendix:selection_function} discusses our approach towards modelling the selection function. We shall show how its impact on our results is minimal.

For the kinematic model, $p(\tilde{\bs{\mu}},s|\ell,b, P)$, we adopt a mixture model that is a combination of an axisymmetric NSD model and a `background' bar/disc model \citep[][hereafter S22]{Sormani2022} -- labelled `bar'. We write
\begin{equation}
\begin{split}
     p(\tilde{\bs{\mu}},s|\ell,b,P) &= \frac{p(\ell,b,\tilde{\bs{\mu}},s|P)}{p(\ell,b|P)}\\&=\frac{w(P) p_\mathrm{NSD}(\ell,b,s,\tilde{\bs{\mu}}|P)+p_\mathrm{bar}(\ell,b,s,\tilde{\bs{\mu}})}{w(P) p_\mathrm{NSD}(\ell,b|P)+p_\mathrm{bar}(\ell,b|P)}.
\end{split}
\end{equation}
$p_\mathrm{NSD}(\ell,b,s,\bs{\mu}|P)$ is the NSD model for each population labelled by period $P$ with $p_\mathrm{NSD}(\ell,b|P)$ its marginalization over distance and proper motion weighted by the selection function, $S$,
\begin{equation}
    p_\mathrm{NSD}(\ell,b|P)\equiv \int \mathrm{d}^2\bs{\mu}\,\mathrm{d}s\, S(\ell,b,s,P) p_\mathrm{NSD}(\ell,b,s,\bs{\mu}|P),
\end{equation}
and likewise for $p_\mathrm{bar}(\ell,b|P)$ (note that although the bar model is independent of period, the marginalized distribution is conditioned on the period as the selection function depends on period). $w(P)$ is a period-dependent weight function that gives the total mass ratio of the NSD stars to the bar stars at fixed period. In later modelling, we either keep $w$ fixed independent of period or allow it to be a flexible interpolated cubic spline with period. 

Our model is not sensitive to the relative number of stars at each $(\ell,b)$ which is dependent on the extinction and detectability of the Mira variables (amongst other things), but is sensitive to the relative fraction, $F(\ell,b, P)$, of stars in each component at each on-sky location $(\ell,b)$ and with a period, $P$:
\begin{equation}
    F(\ell,b,P) \equiv\frac{p_\mathrm{bar}(\ell,b|P)}{w(P) p_\mathrm{NSD}(\ell,b|P)}.
\label{eqn::F_lbP}
\end{equation}
Furthermore, our model is not sensitive to the total normalization (or mass of the model). We opt to normalize $p_\mathrm{NSD}$ by the mass of the NSD found in \citetalias{Sormani2022}, $M_\mathrm{NSD}$. In this way, $w(P)$ is the total mass ratio of the NSD to the bar at fixed period relative to the mass ratio found in \citetalias{Sormani2022} i.e. $w=1$ corresponds to the relative NSD/bar weight found in \citetalias{Sormani2022}. We now describe the specific model components in more detail.

\subsection{Nuclear stellar disc model}
\begin{figure}
    \centering
    \includegraphics[width=\columnwidth]{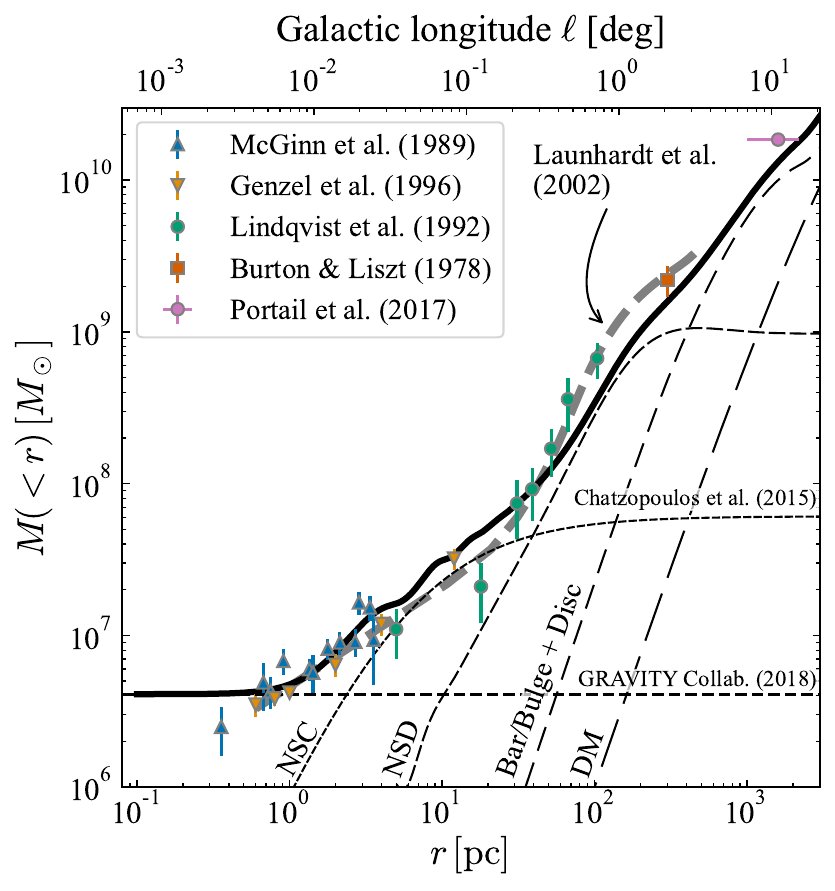}
    \caption{Spherical mass enclosed for the adopted gravitational potential: the solid black line shows the total with the breakdown by different components shown by the series of labelled dashed lines. The thick dashed grey line shows the model of \protect\cite{Launhardt2002}, and the coloured datapoints measurements from the literature (the measurement from \protect\citealt{Portail2017} is given within a cuboid so the depicted errorbar shows the minimum and maximum dimensions of that cuboid). Note Sgr A* is not included in the potential for the modelling in Section~\ref{sec:models}.}
    \label{fig:potential}
\end{figure}

$p_\mathrm{NSD}$ is modelled using action-based distribution functions (DF):
\begin{equation}
    p_\mathrm{NSD}(\ell,b,s,\tilde{\bs{\mu}}|P) = \int\mathrm{d}v_{||}\,s^4\cos b \,f(\bs{J}|P).
\end{equation}
Here $\bs{J}$ are action coordinates computed from a set of observables $(\ell,b,s,\bs{\mu},v_{||})$ with $v_{||}$ the line-of-sight velocity in a choice of axisymmetric potential, $\Phi$ (here fixed). The actions $\bs{J}$ are a triplet of integrals of motion, $(J_r, J_\phi, J_z)$, that approximately give the amplitude of radial oscillation, the degree of circulation and the amplitude of the vertical oscillation of each orbit respectively. We use \textsc{Agama} \citep{Vasiliev2019} for the action computation. $s^4\cos b$ is a Jacobian factor between the observable coordinates and the actions. The potential is modelled by combining the NSC model from \cite{Chatzopoulos2015}, the best-fitting NSD model of \citetalias{Sormani2022} and an axisymmetrized version of the \cite{Portail2017} potential (including the dark matter halo) with the central nuclear component removed as parametrized by \cite{Sormani2022_pot}. The spherical enclosed mass for this potential is shown in Fig.~\ref{fig:potential} alongside mass measurements from \cite{McGinn1989}, \cite{Genzel1996}, \cite{Lindqvist1992}, \cite{BurtonLiszt1978} and \cite{Portail2017}.

$f(\bs{J}|P)$ is the action-based DF parametrized by period. We choose the `quasi-isothermal' class of disc DFs introduced in \cite{Binney2010} and \cite{Binney2011} given by the functional form
\begin{equation}
f(\bs{J}) = \frac{\Omega}{2\pi^2\kappa^2}\tilde\Sigma(R_c)f_R(J_R)f_z(J_z)t(J_\phi)
\end{equation}
where
\begin{equation}
\begin{split}
\tilde\Sigma(R_c) &= \Sigma_0 \exp(-R_c/R_\mathrm{NSD}),\\
f_R(J_R)&=\frac{\kappa}{\tilde\sigma_R^2}\exp(-\kappa J_R/\tilde\sigma^2_R),\\
f_z(J_z)&=\frac{\nu}{\tilde\sigma_z^2}\exp(-\nu J_z/\tilde\sigma^2_z),
\end{split}
\end{equation}
and $t(J_\phi)$ is a taper function given by
\begin{equation}
    t(J_\phi) = \begin{cases}1, & J_\phi\geq0,\\\exp\Big(\frac{2 \Omega J_\phi}{\tilde\sigma_r^2}\Big),&J_\phi<0.
    \end{cases}
\end{equation}
All epicyclic frequencies $(\kappa,\nu,\Omega)$ are evaluated at the radius $R_c$ of a circular orbit with angular momentum
\begin{equation}
\tilde J = \sqrt{(|J_\phi| + k_R J_R + k_z J_z)^2 + \tilde J^2_\mathrm{min}},
\end{equation}
where $k_R=k_z=0.25$ and following \citetalias{Sormani2022} we set $\tilde J_\mathrm{min}=10\,\mathrm{kpc\,km\,s}^{-1}$. The two dispersion functions, $\tilde\sigma_R$ and $\tilde\sigma_z$, are given by
\begin{equation}
\begin{split}
\tilde\sigma_R^2(R_c) &= \sigma^2_{R0} \exp (-2 (R_c-R_\mathrm{NSD})/R_{\sigma,R})+\sigma_\mathrm{min}^2,\\
\tilde\sigma_z^2(R_c) &= 2 H_\mathrm{NSD}^2\nu^2(R_c) + \sigma_\mathrm{min}^2.
\end{split}
\end{equation}
where again following \citetalias{Sormani2022} we set $\sigma_\mathrm{min}=10\,\mathrm{km\,s}^{-1}$.

A disadvantage of the quasi-isothermal DFs is they make explicit reference to the potential through the use of the epicyclic frequencies. This makes them slightly awkward when constructing self-consistent distribution functions. For this reason, alternative DFs have been proposed by \cite{Vasiliev2019} and \cite{Binney2023}. However, we opt to use the quasi-isothermal DFs as (i) they were used in \citetalias{Sormani2022}, (ii) our aim is not to construct self-consistent DFs (the potential is fixed), and (iii) we have found the quasi-isothermal DF parametrization is more physically interpretable than the alternatives.

The introduced DF has four free parameters (the normalization $\tilde\Sigma$ is uninteresting for our purposes): the scalelength $R_\mathrm{NSD}$, the scaleheight $H_\mathrm{NSD}$, the radial dispersion at the scalelength of the disc $\sigma_{R0}$ and the radial scalelength of the dispersion fall-off, $R_{\sigma, R}$ \citepalias[note this is slightly different to][who normalize the radial dispersion at $R=0$]{Sormani2022}. In our modelling, these are either fixed independent of period or modelled as flexible interpolated cubic splines in period.

\citetalias{Sormani2022} used the quasi-isothermal DF models to fit the line-of-sight velocity from \cite{Fritz2021} and proper motion distributions from VIRAC \citep{VIRAC}. They used a fully self-consistent procedure assuming the observed stars traced the underlying mass distribution (modified by a selection function) which together with a fixed potential for the NSC then sourced the potential the DF was computed in. As a reference point for our modelling, we consider the probability $f_{S22}(\bs{J})$ of being part of the NSD as defined by the fitted \citetalias{Sormani2022} distribution function. Note the best-fitting quasi-isothermal DF parameters reported in \citetalias{Sormani2022} do not quite give the best quasi-isothermal DF fit to the NSD in our chosen potential due to the impact of the iterative self-consistency procedure in \citetalias{Sormani2022}. The best-fitting parameters are $(R_\mathrm{NSD}, H_\mathrm{NSD}, \sigma_{R0}, R_{\sigma, R})_\mathrm{S22}=(51\,\mathrm{pc}, 19\,\mathrm{pc}, 112\,\mathrm{km\,s}^{-1}, 330\,\mathrm{pc})$, which will serve as useful data for priors in the subsequent modelling. \citetalias{Sormani2022} found the uncertainty on these parameters was $5$ per cent for $R_\mathrm{NSD}$, $10$ per cent for $H_\mathrm{NSD}$, $5$ per cent for $\sigma_{R0}$ and $R_{\sigma, R}$ is prior-dominated and only a lower limit is obtained. Note that these parameters are only loosely related to the genuine scalelength, scaleheight etc. of the model. For example, the scalelength and scaleheight can be measured directly from a model realisation as $74\,\mathrm{pc}$ and $26\,\mathrm{pc}$ respectively.

\subsection{Bar--disc contamination model}\label{subsection:contaminant_model}
In addition to the component of interest, the NSD, we must also model the contribution from the other foreground and background components of the Galaxy. \citetalias{Sormani2022} has shown that the `contamination' for a spectroscopically targeted sample is significant across the entire NSD region so it is reasonable to assume our Mira variable sample suffers from similar contamination.

The model for this contaminant component, $p_\mathrm{bar}$, is taken from an N-body model from \cite{Portail2017} with the central nuclear component removed (the NSD model fulfils this role in our model). This model has been fitted using the made-to-measure method to bar(-bulge) star counts, spectroscopy and proper motion data and was demonstrated by \citetalias{Sormani2022} to accurately capture the 3D velocity distributions of giant stars in the NSD region when combined with an NSD model. 
We remove model star particles at distances beyond $8\,\mathrm{kpc}$ from the Galactic centre as (i) the outer disc ($R>5.5\,\mathrm{kpc}$) of the \cite{Portail2017} model was imposed as a data-motivated constant-scaleheight exponential disc but not explicitly fitted to data, (ii) it is likely extinction means few very distant disc stars enter our sample and the extinction maps used in Appendix~\ref{appendix:selection_function} only extend to $\sim 10.5\,\mathrm{kpc}$, and most importantly (iii) the Wesenheit cuts employed by \cite{Sanders2022} to clean the sample remove any distant stars \citep[see figure D1 from][]{Sanders2022}.
The \cite{Portail2017} model is rotated such that the major axis of the bar lies at an angle of $25$ deg with respect to the Sun--Galactic Centre line. This component has a fixed functional form in the fitting. Only its relative contribution as a function of $P$ is considered (via $w(P)$).
We compute kernel density estimates (KDE) of $p_\mathrm{bar}(\tilde{\bs{\mu}},s|\ell,b)$ and $p_\mathrm{bar}(s|\ell,b)$ weighted by the mass of each particle for a regular grid of small circular regions in $(\ell,b)$ of solid angle $\Delta\Omega$.
The KDEs are computed using a fast Fourier transform with the KDEpy package\footnote{\url{https://kdepy.readthedocs.io/en/latest/index.html}}.
Furthermore, we store $p_\mathrm{bar}(\ell,b)$ at these locations by simply summing the mass in each region and dividing by the on-sky area of the region. When required, these quantities are linearly interpolated in $(\ell,b)$ for an arbitrary star location. 

\subsection{Coordinate systems}
We follow \citetalias{Sormani2022} and define two Cartesian coordinate systems: one centred on $(\ell,b)=(0,0)$ with the Sun-Galactic centre line in the $(x,y)$ plane and the Galactic Centre a distance $8.275\,\mathrm{kpc}$ from the Sun \citep{GravityCollaboration}, and a second aligned with the first but shifted by $8.0\,\mathrm{pc}$ in the direction opposite Galactic rotation and $6.6\,\mathrm{pc}$ towards the Galactic South Pole such that the origin of the system is centred on Sgr A* at $(\ell,b)=(-0.056,-0.046)\,\mathrm{deg}$. We set the solar motion as $4.74\times 6.41\,\mathrm{mas\,yr}^{-1} \times 8.275\,\mathrm{kpc}=251.4\,\mathrm{km\,s}^{-1}$ in the $\ell$ direction \citep{Reid2020}, $11.1\,\mathrm{km\,s}^{-1}$ towards the centre of the Galaxy and $7.25\,\mathrm{km\,s}^{-1}$ towards the Galactic North Pole \citep{Schoenrich2010}. The multi-component potential and action coordinates are computed in the frame centred on Sgr A*.

\subsection{Computational specifics}\label{sec::comp_specifics}
Our fitting procedure relies on marginal distributions so per-star integrals are required. We follow \cite{ZhangSanders2023} in the efficient computation of these integrals. $N_s$ proper motion samples are generated for each star from the uncertainty distributions: $\bs{\mu}'\sim\mathcal{N}(\tilde{\bs{\mu}},\bs{\Sigma_\mu})$\footnote{We use the notation $x\sim\mathcal{N}(\mu,\sigma)$ to denote that $x$ is a random variate drawn from a normal distribution with mean $\mu$ and standard deviation $\sigma$, and $\bs{x}\sim\mathcal{N}(\bs{\mu},\bs{\Sigma})$ for the vector version where $\bs{\Sigma}$ is a covariance matrix. $\mathcal{N}(x|\mu,\sigma)$ denotes evaluating the normal distribution with mean $\mu$ and standard deviation $\sigma$ at $x$, and similar for the vector version.}. These are complemented by line-of-sight velocity samples and distance samples drawn from two Gaussians: $v'\sim\mathcal{N}(v_0,\sigma_v)$ with $v_0=0$ and $\sigma_v=200\,\mathrm{km\,s}^{-1}$ and $s'\sim\mathcal{N}(s_0,\sigma_s)$ with $s_0=8.275\,\mathrm{kpc}$ and $\sigma_s=0.2\,\mathrm{kpc}$ respectively. The actions and frequencies for each star's set of samples are pre-computed. Then, for each star, we compute $p_\mathrm{NSD}(\ell,b,s,\tilde{\bs{\mu}}|P)$ as
\begin{equation}
p_\mathrm{NSD}(\ell,b,\tilde{\bs{\mu}}|P)\approx \frac{1}{N_s} \sum_i^{N_s} \frac{S(\ell,b,s_i',P)\, s_i'^4\cos b \, f(\bs{J}')}{\mathcal{N}(v'|v_0,\sigma_v)\mathcal{N}(s'|s_0,\sigma_s)}.
\end{equation}
Similarly, the denominator terms are found by drawing samples for $\bs{v}'=(v_\ell',v_b',v_\mathrm{los}')\sim \mathcal{N}(\bs{v}_0,\mathrm{diag}(\bs{\sigma}_v))$ with $\bs{v}_0=(-250,0,0)\,\mathrm{km\,s}^{-1}$ and $\bs{\sigma}_v=(100,100,100)\,\mathrm{km\,s}^{-1}$ pre-computing all of the actions and frequencies ($\bs{J}'$), and evaluating
\begin{equation}
p_\mathrm{NSD}(\ell,b|P)\approx \frac{1}{N_s} \sum_i^{N_s} \frac{S(\ell,b,s_i',P)\, s_i'^2\cos b \, f(\bs{J}')}{\mathcal{N}(\bs{v}'|\bs{v}_0,\bs{\sigma}_v)\mathcal{N}(s'|s_0,\sigma_s)}.
\end{equation}
Note that as we are generating samples in the physical velocity space, not the proper motion space the Jacobian is $s^2$ instead of $s^4$. We use $N_s=500$ samples for the `numerator' quantities and $N_s=1000$ for the `denominator'. We adopt a very similar procedure for the computation of the integrals over the background bar model as
\begin{equation}
    p_\mathrm{bar}(\ell,b,\tilde{\bs{\mu}}|P) \approx \frac{1}{N_s'}\sum_i^{N_s'}\frac{S(\ell,b,s_i',P) p_\mathrm{bar}(s,\tilde{\bs{\mu}}|\ell,b)p_\mathrm{bar}(\ell,b)}{\mathcal{N}(s'|s_0,\sigma_s')},
\end{equation}
where $N_s'=10N_s$, $\sigma_s'=3\,\mathrm{kpc}$ and the terms in the numerator are described in Section~\ref{subsection:contaminant_model}. The computation of $p_\mathrm{bar}(\ell,b|P)$ is very similar (using $p_\mathrm{bar}(s|\ell,b)$ instead of $p_\mathrm{bar}(s,\tilde{\bs{\mu}}|\ell,b)$). Where possible, these quantities are pre-computed. With all components of our modelling approach defined, we now turn to fitting the models to the data sample.

\subsection{Priors and model implementation}\label{sec::priors}
We place priors on the logarithms of the various parameters to ensure positivity. When considering $w(P)$ independent of period, we adopt the prior $\ln w \sim \mathcal{N}(0,3)$. For more flexible models we allow the relative weight $\ln w(P)$ and the NSD DF parameters $(\ln R_\mathrm{NSD}, \ln H_\mathrm{NSD}, \ln \sigma_{R,0}, \ln R_{\sigma, R})$ to be fitted interpolated cubic splines in $\log_{10}P$. By default, we use $12$ equally-spaced knots in $\log_{10}P/\mathrm{d}$ between $1.9$ and $3.1$ (although we also consider equally-spaced in age later). The $12$ knot values for each free function are the parameters of the model on which we place priors. Extreme regions of parameter space can be assigned high likelihood due to poor estimates of the integrals from the samples we use. For this reason, we introduce smooth lower limits to the parameters at $(R_\mathrm{NSD},  H_\mathrm{NSD}, \sigma_{R,0}, R_{\sigma,R}, 1/w(P))=(5\,\mathrm{pc},5\,\mathrm{pc},50\,\mathrm{km\,s}^{-1},50\,\mathrm{pc},10^{-6})$.

Identifiability of the model components is important. Giving the NSD component too much freedom allows it to replicate the bar-bulge component and we are unable to distinguish NSD vs. bar-bulge stars. From the work of \citetalias{Sormani2022}, we have prior knowledge of the approximate form of the NSD DF. We, therefore, place priors on all the knots as
\begin{equation}
    \begin{split}
        \ln w(P) &\sim \mathcal{N}(0,3),\\\ln R_{\mathrm{NSD}} &\sim \mathcal{N}(\ln R_{\mathrm{NSD},\mathrm{S22}},0.6),\\\ln H_{\mathrm{NSD}} &\sim \mathcal{N}(\ln H_{\mathrm{NSD},\mathrm{S22}},0.6),\\\ln \sigma_{R,0} &\sim \mathcal{N}(\ln \sigma_{R,0,\mathrm{S22}},0.3),\\\ln R_{\sigma,R} &\sim \mathcal{N}(\ln R_{\sigma,R,\mathrm{S22}},0.3),
    \end{split}
\label{eqn:priors}
\end{equation}
and then we introduce a smoothing prior $P_\mathrm{smooth}$ e.g. 
\begin{equation}
    \ln P_\mathrm{smooth} \propto \sum_i -\frac{(\ln w_i(P)-\ln w_{i+1}(P))^2}{2\ln \sigma_w^2}-\ln\ln \sigma_{w}
\end{equation} for $w(P)$ where the smoothing scales are hyperparameters following half-normal priors $\sigma_i\sim\mathcal{N}(0,\tau_i)$ where 
\begin{equation*}
    \ln (\tau_{R_\mathrm{NSD}},\tau_{H_\mathrm{NSD}},\tau_{\sigma_{R,0}},\tau_{R_{\sigma,R}},\tau_w)=(0.3,0.3,0.3,0.3,1).
\end{equation*}
Note that this combination of a Gaussian prior on each knot value and a smoothing prior makes the effective prior on each knot value tighter than just the Gaussian prior would suggest.
$12$ knots is a good compromise between speed and flexibility. As smoothing priors are adopted, the risk of `over-fitting' is minimal and more knots are preferable to capture all significant data features. To check the results, we also fit models binned by period adopting the priors given in equation~\eqref{eqn:priors}. The models are implemented in \textsc{Jax} \citep{jax} and \textsc{Numpyro} \citep{numpyro1,numpyro2}, and sampled using the NUTS sampler \citep{nuts_sampler}.

\section{Dynamical model fitting results}\label{sec:results}

\begin{figure*}
\centering
    \includegraphics[width=\textwidth]{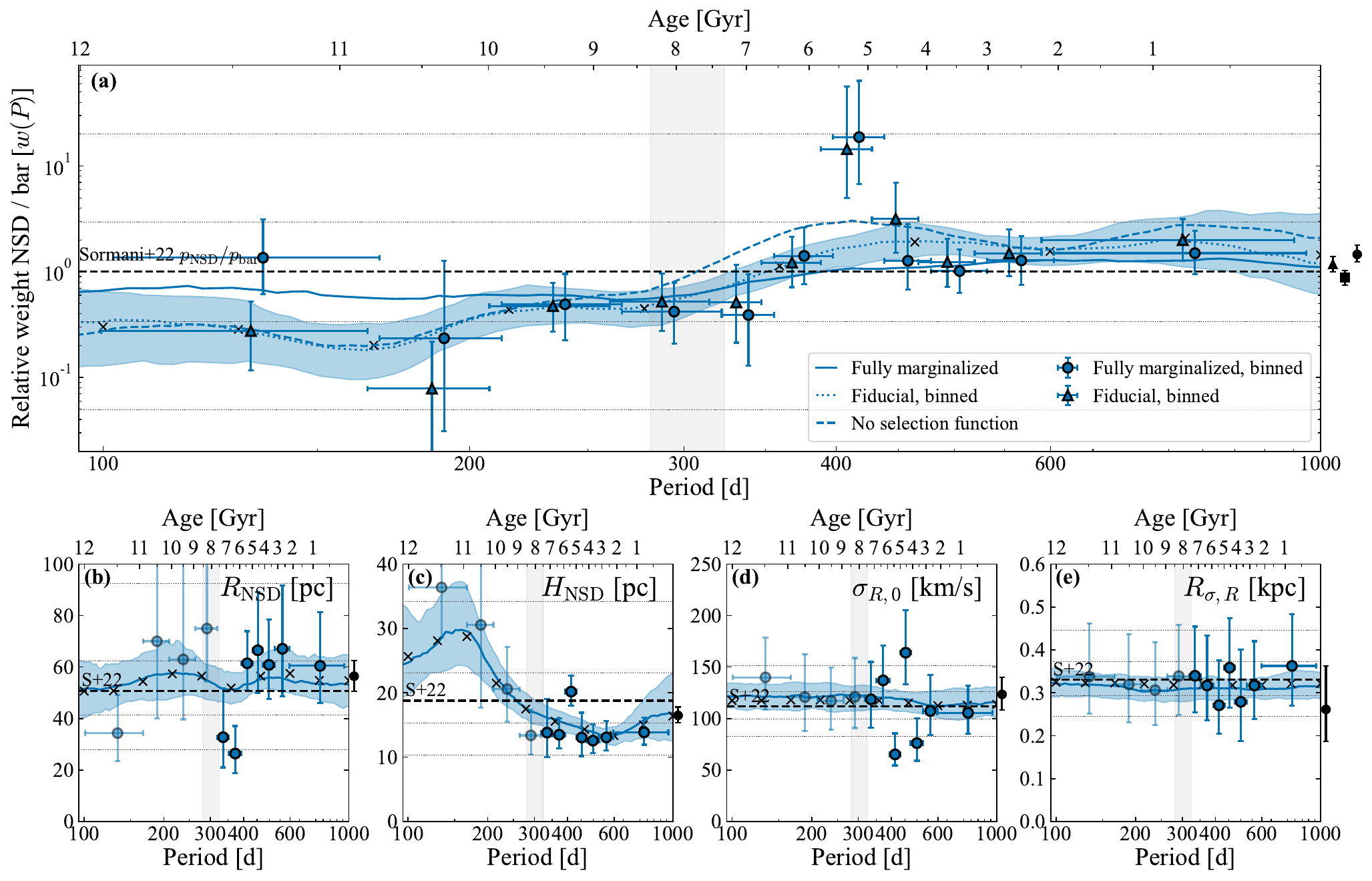}
    \caption{Dynamical model fitting results: the large panel (a) shows the relative weight of the NSD to the bar ($w(P)$ in the modelling). The points are estimates using Mira variables split into bins (circles are the fully marginalized model and triangles the model with a fixed DF from \protect\cite{Sormani2022}). The three lines are from the spline models with fixed weights (solid), no selection function (dashed) and fully marginalized (dotted). The bottom row of panels (b--e) shows the NSD DF fit results for the fully-marginalized model as a function of period (age) (the scalelength, $R_\mathrm{NSD}$ in panel (b), scaleheight $H_\mathrm{NSD}$ in panel (c), central radial dispersion $\sigma_{R,0}$ in panel (d) and the radial scalelength of the radial dispersion fall-off $R_{\sigma,R}$ in panel (e). The binned fits are the points and the line is the spline fit result. Faint horizontal lines show the $\pm 1\sigma$ for the prior width for the spline (inner) and binned (outer) models. The black error bars outside each panel show the results for the period-independent models (square: fiducial, circle: fully marginalized, square: no selection function).}
    \label{fig:parameters}
\end{figure*}

We use the same sample of $1392$ stars described in Section~\ref{sec:data}: `reliable', high-amplitude ($\Delta K_s>0.4$), low latitude ($|b|<0.4\,\mathrm{deg}$) and with $>3\sigma$ outliers in proper motion removed.

\subsection{Global models}
\begin{table}
    \caption{Fitted DF parameters for models independent of period. The first columns give the results fixing the NSD DF parameters to their values from \protect\cite{Sormani2022} with and without a selection function. The third column gives the results when the DF parameters are also fitted incorporating priors given in the fourth column (imposed in log space hence the slight inconsistency for $w$).}
    \centering
    \begin{tabular}{l|cccc}
Parameter & Fixed & No SF & Marg. & Prior (\citetalias{Sormani2022})\\
\hline
$w$ & 	$0.8 \pm 0.1$&$1.1 \pm 0.2$&$0.7 \pm 0.1 $&$ 1.0 \pm 10.0$\\
$R_\mathrm{NSD}$ [pc] & 	-&-&$56 \pm 6 $&$ 51 \pm 32$\\
$H_\mathrm{NSD}$ [pc] & 	-&-&$17 \pm 1 $&$ 19 \pm 12$\\
$\sigma_{R,0}$ [km/s] & 	-&-&$123 \pm 16 $&$ 112 \pm 34$\\
$R_{\sigma,R}$ [pc] & 	-&-&$261 \pm 88 $&$ 331 \pm 101$\\
    \end{tabular}
    \label{tab:param_results}
\end{table}
We begin by performing `global fits' which neglect any dependence of the NSD properties with period. The results of these investigations are shown in Table~\ref{tab:param_results}. 

First, we consider a model where the relative weight $w$ is independent of period. The uncertainty on the mass from the \citetalias{Sormani2022} model is around $10$ per cent and the bar-disc total mass within the region is uncertain to around $6$ per cent \citep{Portail2017} such that from prior data $w$ is constrained as $w=(1.00\pm0.12)$. We use the NSD weights from the \citetalias{Sormani2022} model, $f_\mathrm{S22}(\bs{J})$, finding $w=(0.85\pm 0.14)$. Excluding the selection function leads to $w=(1.14\pm 0.18)$ as the selection function biases towards the more distant, background stars which mimic the colder dispersion of the NSD meaning a lower weight of NSD is required. However, in both cases $w$ is consistent with the unity value found by \citetalias{Sormani2022} demonstrating that the sample is representative of stars in the NSD region. If we simultaneously fit the DF parameters alongside the relative weight using the priors in equation~\eqref{eqn:priors}, we find NSD parameters highly consistent with the results of \citetalias{Sormani2022} (see Table~\ref{tab:param_results}) although our constraint on $R_{\sigma,R}$ is entirely prior-dominated. In this model, the relative weight increases to $w=(0.69\pm0.13)$ but is still consistent with the measurement from \citetalias{Sormani2022} within the respective uncertainties. These models further confirm that our sample contains NSD and bar-bulge stars and is relatively unbiased (when accounting for the selection function).

\subsection{Period-dependent models}
We now model the variation of the weighting term, $w(P)$, with period. As described in the previous section, we use an interpolated cubic spline. The DF parameters are fixed to their values from \citetalias{Sormani2022}. We refer to this model as our `\emph{Fiducial}' model. We show the resulting fit in Fig.~\ref{fig:parameters}. We notice that at short period the relative NSD to bar weight is significantly smaller than $w=1$ (the average value over the full population) whilst at long period the NSD dominates with $w\approx2$. The transition occurs around a period of $350$ days \citep[corresponding to an approximate age of $7\,\mathrm{Gyr}$ using the relations from][]{ZhangSanders2023}. This transition appears suggestive of the formation of the NSD but we delay the detailed consequences of this model to a later section.

We further fit a model allowing all the NSD DF parameters to be free functions of $P$, dubbed `\emph{Fully marginalized}' model, using the priors given in Section~\ref{sec::priors}. As shown in Fig.~\ref{fig:parameters}, we find a much weaker transition from NSD kinematics to bar kinematics, but still the transition occurs around the same period. As discussed previously, identifiability becomes an issue and a very thick NSD can resemble the bar population. Indeed that appears to be happening as seen in the lower panels of Fig.~\ref{fig:parameters} where panel (c) echoes the conclusion of Fig.~\ref{fig:mix}. The other parameters are relatively flat with period. We observe that periods longer than $300\,\mathrm{day}$ (so ages less than $\sim8\,\mathrm{Gyr}$) there is a weak suggestion of inside-out formation ($R_\mathrm{NSD}$ decreasing with age) and dynamical heating ($H_\mathrm{NSD}$ increasing with age). However, largely the data appears consistent with no gradients in period. This is somewhat at odds with the results from \cite{NoguerasLara2022} for the Milky Way and \cite{Bittner2020} for external galaxies. However, dynamical mixing is likely significant in the NSD (where the orbital time is $\sim1\,\mathrm{Myr}$) so any formation gradients may get rapidly washed out \citep[e.g.][]{Frankel2020}. These conclusions are corroborated by the fits binned by period also shown in Fig.~\ref{fig:parameters}. In the bin around $\sim400$ days, a very large NSD fraction is found, possibly suggesting a very significant burst at this epoch that is smoothed over by the spline model, but the uncertainties are large.

\subsection{Model variants}
We now explore a number of model variants that test different assumptions in our modelling. 
Although our selection function is well motivated, it is quite likely that there are some shortcomings in the approach. As an alternate extreme, we now consider a model \emph{without} a selection function term (i.e. $S=1$), dubbed `\emph{No selection function}' model. We have already seen in the `global model' that the selection function has minimal impact. Here we opt to fix the DF parameters with period to the \citetalias{Sormani2022} values. We display the result in the top panel of Fig.~\ref{fig:parameters} which agrees well with the fiducial model incorporating the selection function but with a slightly shorter period transition from bar-dominated to NSD-dominated kinematics.

One concern with the modelling approach is that the 3D extinction maps are not sufficient to resolve extinction within the NSD region. It could well be that in some regions we fail to observe Mira variables on the far side of the NSD due to extreme extinction inside the NSD. This effect will severely bias the Galactic longitude proper motion distributions effectively removing one of the velocity peaks. However, the Galactic latitude proper motion distributions will be less biased as at the distance of the NSD the near- and far-sides of the NSD have effectively the same distributions. We, therefore, run a model, dubbed the `\emph{Only using $\mu_b$}' model, ignoring the $\mu_\ell$ measurements and instead marginalize over them in the way described in Section~\ref{sec::comp_specifics}. This still ignores any extreme extinction variation in the NSD but makes the modelling less sensitive to such effects. Finally, as highlighted in Section~\ref{sec:data}, the proper motion uncertainties may be poorly estimated both from saturation effects (see Fig.~\ref{fig:uncert}) and also more broadly due to other calibration issues \citep{Luna2023}. We, therefore, consider the proper motion uncertainties inflated by the unit weight error shown in Fig.~\ref{fig:uncert} and broadened by a further factor of $1.1$ \citep{Luna2023}. This model is dubbed the `\emph{Scaled proper motion errors}' model. Both the `\emph{Only using $\mu_b$}' model and the `\emph{Scaled proper motion errors}' model produce very similar results to the fiducial model as we will see in the next section.

\subsection{Posterior predictive checks from mock sample generation}
\begin{figure*}
\centering
    \includegraphics[width=0.98\textwidth]{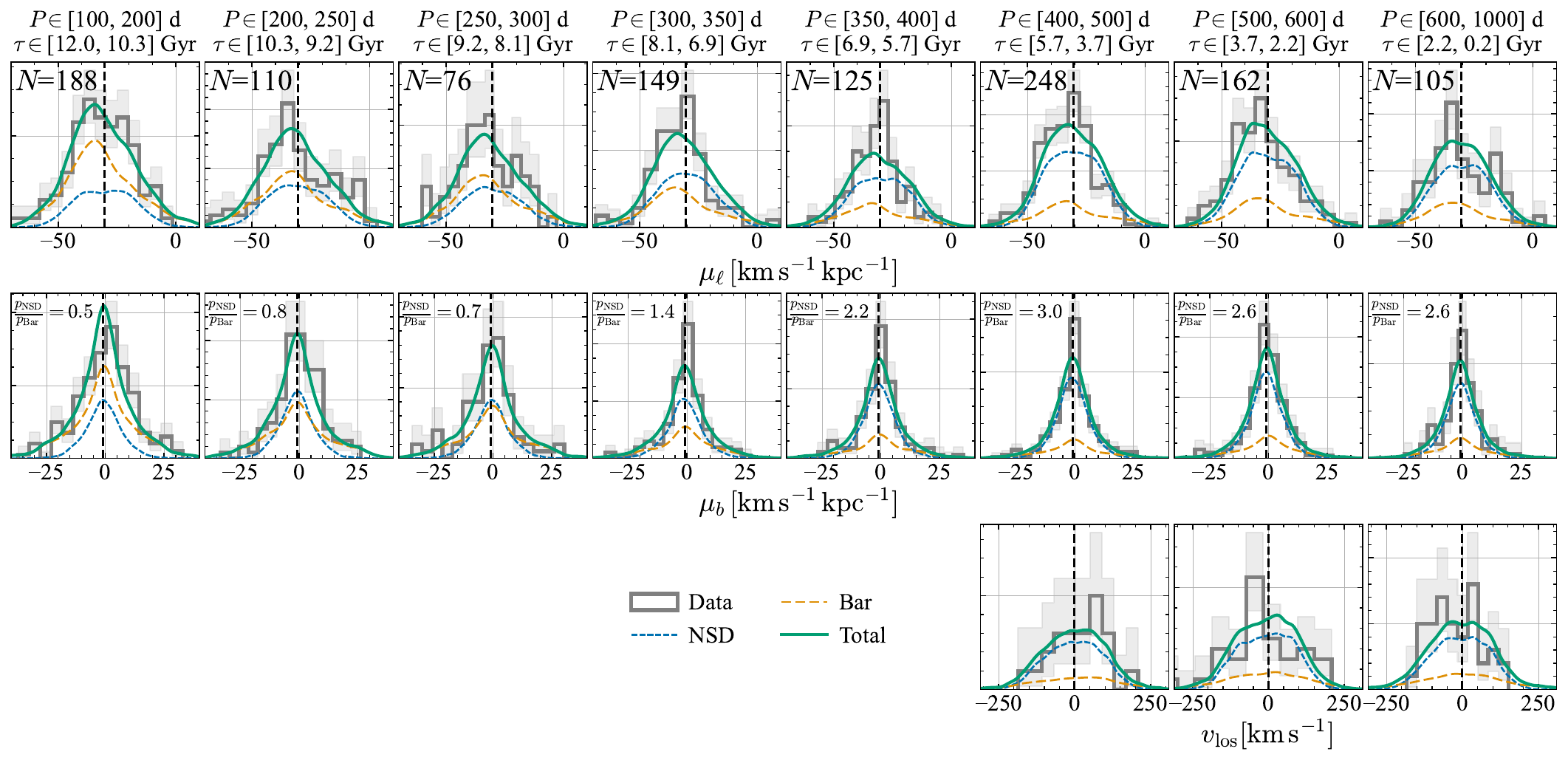}
    \caption{Velocity distributions for the data (grey with Poisson uncertainty shaded interval) compared to the best-fitting model split by period (age).  The green line is the total model, blue short-dashed the NSD model and orange long-dashed the bar/disc contamination model. The top row shows the Galactic longitude proper motions, $\mu_\ell$, the middle row the Galactic latitude proper motions $\mu_b$ and the bottom row line-of-sight velocities, $v_\mathrm{los}$, not used in the analysis. Each column corresponds to a different period bin as indicated above the top row. In the top row, the number of stars in the proper motion panels is given and in the middle panel, the model fraction of the NSD to the bar is shown.}
    \label{fig:veldists}
\end{figure*}

To perform posterior predictive checks, we generate mock samples from the fitted models. We use \textsc{emcee} \citep{ForemanMackey2017} to generate samples in the observable space $(\ell,b,s,\bs{\mu},v_{||},P)$ from $s^4\cos b\, f(\bs{J}|P)\mathcal{U}(\log_{10}P)$ where $\mathcal{U}$ is a uniform distribution and assign a total mass of $M_\mathrm{NSD}$ to the samples. We then select a subset of particles with associated masses from the \cite{Portail2017} models within the on-sky NSD region and assign them periods log-uniformly sampled between $100$ and $1000$ days. We multiply the NSD particle masses by $w(P)$ and then all particle masses by the selection fraction $S(\ell,b,s,P)$. From the combined set of NSD and bar samples, we randomly draw a subset of particles with probability proportional to their particle mass. This sample has the correct balance of bar and NSD at each location but does not capture the on-sky and period distributions of the data. To do this, we, for each datum, find the $100$ nearest mock stars in $(\ell,b)$ and $P$ using a Euclidean distance with scales of $\Delta \log_{10}P=0.05$ and $(\Delta\ell,\Delta b)=(0.05,0.05)\,\mathrm{deg}$. Finally, we convolve the mock proper motions with the errors of the matched datum. When a sample unaffected by selection function effects is required, we do not multiply the masses by the selection function and instead compute the period distribution of the data using the procedure described in the next subsection, and then reweight the samples by this distribution.

The results of this mock generation procedure for the fiducial model are shown in Figs.~\ref{fig:veldists} and~\ref{fig:veldists_onsky} in Appendix~\ref{sec:additional_posterior} where the proper motion and line-of-sight velocity histograms for the data are compared to the model samples. From Fig.~\ref{fig:veldists}, we observe, for all period bins, a good model fit to the data. The line-of-sight velocities are not used in the fits so they give good corroboration of the results and demonstrate the power of using a dynamical model. As the period increases, there is an increasing dominance of the NSD relative to the bar.

\begin{figure}
    \centering
    \includegraphics[width=\columnwidth]{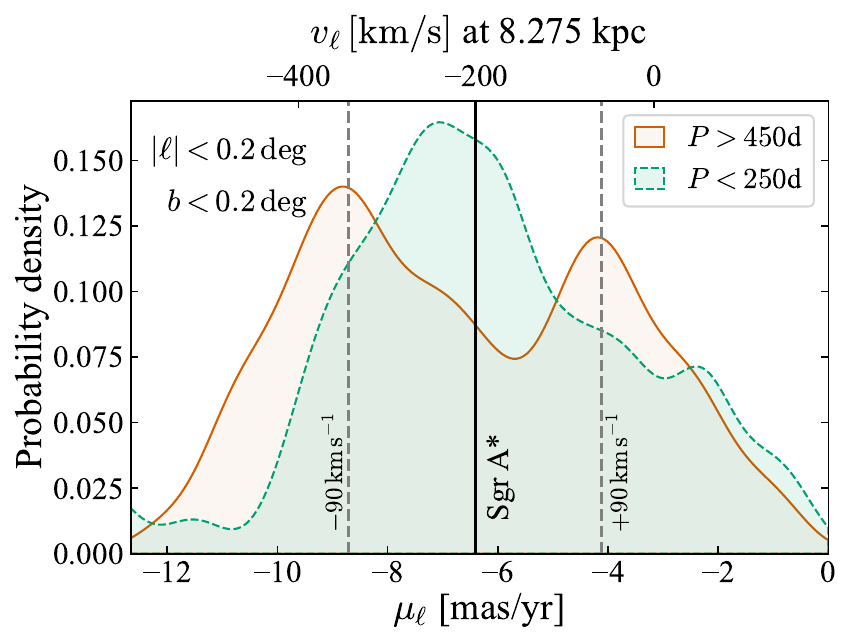}
    \caption{Galactic longitude proper motion distributions for two samples located $|\ell|<0.2\,\mathrm{deg}$ and $b<0.2\,\mathrm{deg}$ and with proper motion uncertainties $<1.5\,\mathrm{mas\,yr}^{-1}$. The blue solid line shows a `young' sample with periods $>450$ day where the rotation of the NSD is visible whilst the orange dashed line shows an `old' sample with periods $<250$ day showing no net rotation with respect to Sgr A*.}
    \label{fig:kinematic_split}
\end{figure}

Further evidence of the presence of NSD member stars in our sample at long periods is given in Fig.~\ref{fig:kinematic_split}. By restricting to stars in a $0.4\,\mathrm{deg}\times0.4\,\mathrm{deg}$ region centred on $(\ell,b)=(0,0)$, we see that short-period ($P<250$ d, old) stars have a unimodal $\mu_\ell$ distribution centred on the motion of Sgr A* with a skew towards positive $\mu_\ell$ characteristic of the background bar-disc population, whilst the long-period ($P>450$ d, young) stars have a bimodal distribution characteristic of the rotating NSD population \citep[e.g.][]{Shahzamanian2021} with a rotation amplitude of $\sim90\,\mathrm{km\,s}^{-1}$. This demonstrates that our sample appears to probe both sides of the NSD and we are not limited by dust within the NSD region (at least in some parts).

\subsection{NSD star formation history and the bar formation time}
\begin{figure}
    \centering
    \includegraphics[width=\columnwidth]{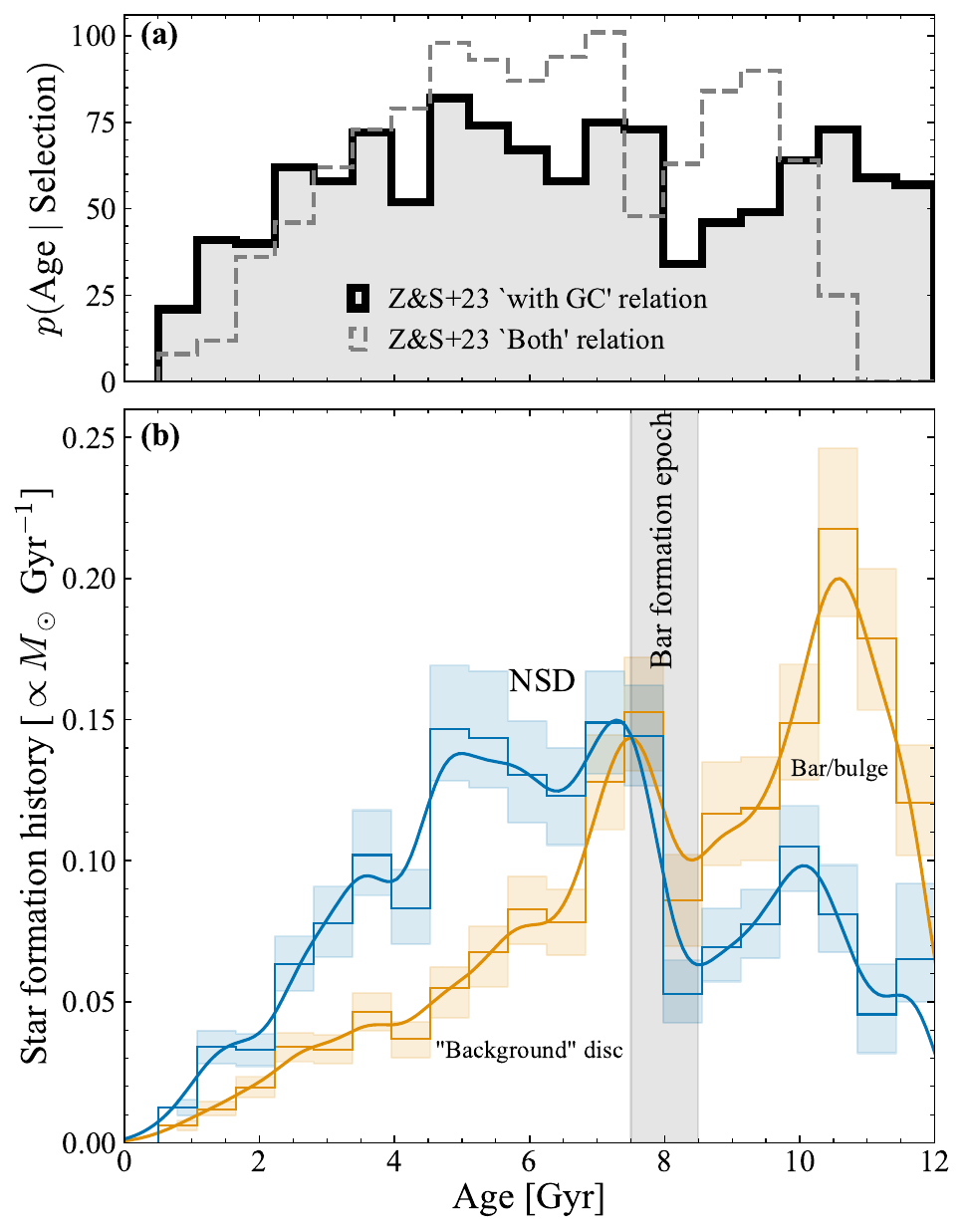}
    \caption{Star formation history of the NSD: \textbf{panel (a)} shows the `raw' age distribution of the sample using two period--age relations as described in the text. \textbf{Panel (b)} shows the resulting inferred \emph{normalized} star formation histories for the NSD (blue) and the bar/disc background (orange).}
    \label{fig:star_formation_history}
\end{figure}

\begin{figure*}
    \centering
    \includegraphics[width=\textwidth]{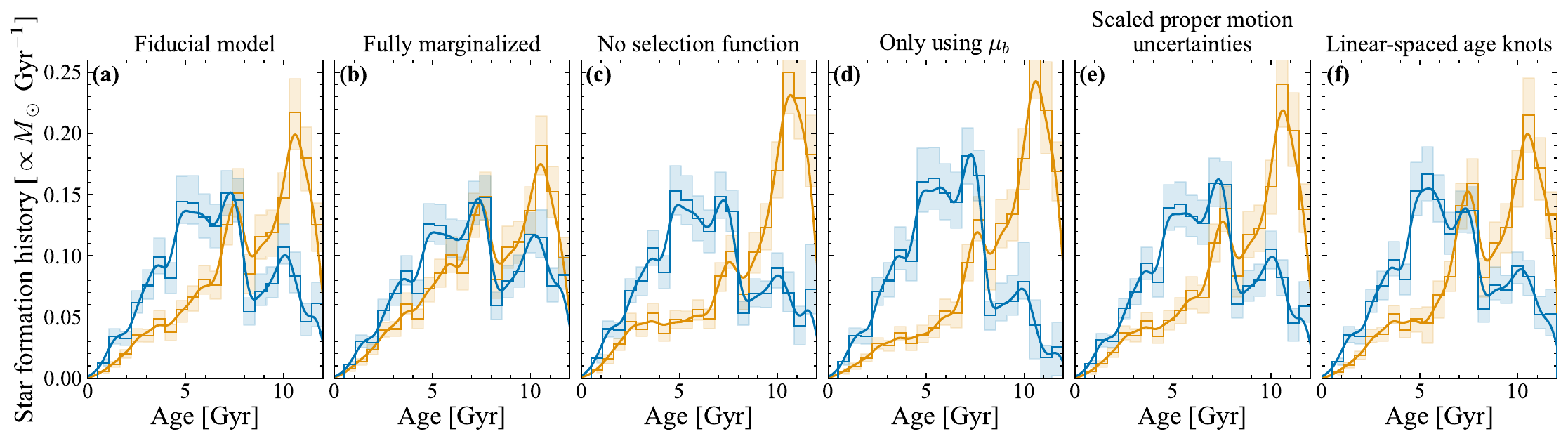}
    \caption{Inferred star formation histories with differing modelling assumptions (as denoted above each panel). See caption of Fig.~\ref{fig:star_formation_history} for more details.}
    \label{fig:star_formation_history_comparison}
\end{figure*}
We now proceed to estimate the star formation history of the NSD and in turn the epoch of formation of the Galactic bar. Under the assumptions of our model, the relative contribution of NSD stars and contaminant bar/disc stars at each location is given by $F(\ell,b,P)$ from equation~\eqref{eqn::F_lbP}. Therefore, we can convert the observed period distribution at a given location $p(P|\ell,b)$ into the period distribution of the NSD, $p_\mathrm{NSD}(P)$, as
\begin{equation}
    p_\mathrm{NSD}(P) = \frac{p_{\mathrm{NSD},S=1}(\ell,b|P)}{p_{\mathrm{NSD}}(\ell,b|P)}\frac{p(P|\ell, b)}{1+F(\ell,b,P)},
\end{equation}
and similarly using the weight factor $F/(1+F)$ for $p_\mathrm{bar}(P)$. The first ratio is the impact of the selection function on the period distributions: the ratio of the on-sky density given period under the selection function divided by the same but without considering the selection function. The second ratio gives the observed period distribution of the NSD which is affected by the selection function. Multiplication by the first term undoes this effect. The left-hand side is independent of the on-sky position so can be estimated using any subset of stars. We, therefore, construct histograms of the period distribution of the sample weighted by $(1+F)^{-1}$ and have confirmed that similar histograms are obtained when limiting only to `high'-latitude stars. To convert these distributions into star formation histories we have to first adopt a period--age relation. We consider the `Both' and `With GC'\footnote{GC for globular cluster not to be confused with GC for Galactic Centre.} relations from table 3 of \cite{ZhangSanders2023}. These share the parametric form
\begin{equation}
\tau(P) = \frac{\tau_0}{2}\Big(1+\tanh\Big[\frac{330\,\mathrm{d}-P}{P_0}\Big]\Big),
\end{equation}
with $(\tau_0,P_0)=(13.7\,\mathrm{Gyr}, 401\,\mathrm{d})$ for the `Both' fit and $(\tau_0,P_0)=(14.7\,\mathrm{Gyr}, 308\,\mathrm{d})$ for the `With GC' fit.
The first of these fits is a joint fit to the period--velocity dispersion in both the radial and vertical directions, whilst the second also includes information from cluster members. Both relations are quite similar but the `With GC' relation is slightly steeper. They are also consistent with other relations used in the literature \citep[e.g.][]{WyattCahn1983, FeastWhitelock1987, FeastWhitelock2014, Catchpole2016, LopezCorredoira2017, Grady2020, Nikzat2022}. There is likely significant scatter in the Mira period--age relation as demonstrated in the theoretical models from \cite{Trabucchi2019}. Based on the differences between dynamical ages and cluster member ages, \cite{ZhangSanders2023} quote a relative age scatter at fixed period in their `With GC' fit of $11$ per cent.

The next step is to convert the number density in age into a star formation rate. The number of Mira variables per unit age, $\tau$, given a star formation rate law $\Gamma(\tau)\equiv\mathrm{d}M_\star/\mathrm{d}\tau$ and an IMF $\xi(M)\equiv\mathrm{d}N/\mathrm{d}M=(M/M_\odot)^{-2.3}/\mathcal{N}$ (the number of stars formed per unit stellar mass normalized to give a total mass of star formation of $M_\star=M_\odot$) is
\begin{equation}
    \frac{\mathrm{d}N}{\mathrm{d}\tau} = \int_{M_\mathrm{l}(\tau)}^{M_\mathrm{u}(\tau)} \mathrm{d}M\, \Gamma(\tau) \xi(M),
\end{equation}
where $M_\mathrm{u}(\tau)$ and $M_\mathrm{l}(\tau)$ are the upper and lower masses of stars of age $\tau$ in the Mira phase. If the Mira phase lasts $\tau_\mathrm{Mira}(M)$ and a star becomes a Mira variable after approximately the main sequence lifetime of $\tau = (10\,\mathrm{Gyr})(M/M_\odot)^{-2.5}$ then we can write
\begin{equation}
\begin{split}
    \frac{\mathrm{d}N}{\mathrm{d}\tau} &\approx \tau_\mathrm{Mira}(\tau) \Big(-\frac{\mathrm{d}M}{\mathrm{d}\tau}\Big)  \Gamma(\tau) \xi(M),
    \\&=\frac{0.04\,\mathrm{Gyr}^{-1}}{\mathcal{N}}\tau_\mathrm{Mira}(\tau) \Big(\frac{10\,\mathrm{Gyr}}{\tau}\Big)^{0.48}  \Gamma(\tau).
\end{split}
\end{equation}
The $\tau^{0.48}$ factor gives the combination of the number of stars formed of age $\tau$ that have masses consistent with being giant stars and the rate at which these giant stars are forming. $\tau_\mathrm{Mira}(\tau)$ gives the time stars born an age $\tau$ ago will be a Mira variable for. This quantity is quite uncertain. \cite{Trabucchi2019} presented models showing the time spent in the Mira fundamental pulsation phase for stars of metallicity $Z=0.008$ and masses $M=(1.5,2.6,4.8)M_\odot$ is $\sim(0.2,0.15,0.3)\,\mathrm{Myr}$. This suggests we can assume $\tau_\mathrm{Mira}$ is approximately constant with age. In conclusion, we can map the period distribution to a star formation rate using
\begin{equation}
    \frac{\mathrm{d}M_\star}{\mathrm{d}\tau} \propto \tau^{0.48}  \frac{\mathrm{d}P}{\mathrm{d}\tau}\frac{\mathrm{d}N}{\mathrm{d}P}.
\end{equation}

\begin{table}
    \caption{Bar formation epoch estimates: best estimates of the bar age using different modelling configurations. The scatter and possible systematic error in the period--age relation is around $1\,\mathrm{Gyr}$ as is the typical time difference between bar formation and the peak star formation in the NSD. We quote $3$ significant figures just to illustrate the differences between the models.}
    \centering
    \begin{tabular}{l|c}
        Setup & Estimate [Gyr]\\
        \hline
Fiducial&$7.89$\\
Alternate period--age relation&$7.29$\\
Fully marginalized&$7.89$\\
No selection function&$7.88$\\
Only using $\mu_b$&$7.87$\\
Scaled proper motion errors&$7.89$\\
Linearly-spaced age knots&$7.91$
    \end{tabular}
    \label{tab:bar_age}
\end{table}

We apply this procedure using the `fiducial' model as displayed in Fig.~\ref{fig:star_formation_history}. The top panel of this figure shows the `raw' age distribution using both period--age relations from \cite{ZhangSanders2023}. In the lower panel, we show the inferred star formation history of the NSD and the bar contaminant model. We define the `bar formation epoch' as the time at which the NSD star formation history is maximally increasing. \cite{Baba2019} show from simulations that the NSD star formation history has a peak after bar formation validating our choice. From Fig.~\ref{fig:star_formation_history} we see this corresponds to $\sim8\,\mathrm{Gyr}$ for the fiducial model using the `with GC' period--age relation from \cite{ZhangSanders2023}. The full set of results is given for each of the model variants in Table~\ref{tab:bar_age}. We see that all models are consistent with the $\sim8\,\mathrm{Gyr}$ bar age estimate. Using the period--age relation from \cite{ZhangSanders2023} that doesn't consider the cluster results (`Both') produces a $0.6\,\mathrm{Gyr}$ lower bar age estimate so we can consider an approximate $1\,\mathrm{Gyr}$ systematic uncertainty in the bar age estimate arising from the choice of period--age relation. \cite{ZhangSanders2023} also investigated biases arising from velocity uncertainties in the age-velocity-dispersion relation calibration. By assuming a uniform stellar age distribution they found that the period--age relation is essentially insensitive to velocity uncertainties around $8\,\mathrm{Gyr}$ meaning the systematic uncertainty on our result from this source of calibration error is subdominant. 
Fig.~\ref{fig:star_formation_history} also supports the metric proposed by \cite{Baba2022} where in a simulation without bar buckling the bar formation time is estimated from the transition age between the bar-bulge star formation history and that of the NSD. In a similar vein, in Appendix~\ref{sec:timer} we consider how the Milky Way bar age would be inferred by an external observer of the galaxy using the integrated field unit method of \cite{deSaFreitas2023, deSaFreitas2023b} finding consistent estimates. This gives good corroboration that methods for external galaxies match expectations when applied to observational data.

\section{Discussion and Conclusions}\label{sec:discussion}
In summary, our results from analysing the proper motions of Mira variables in the NSD region indicate:
\begin{enumerate}
    \item Around the $350$ day period there is a transition in the proper motion kinematics with the shorter period (older) Mira variables appearing to have more bar-like kinematics whilst the longer period (younger) have more NSD-like kinematics (both colder kinematics and a multi-modal rotation signature). This is evidenced by detailed distribution function modelling accounting for selection effects (Fig.~\ref{fig:star_formation_history}), more basic Gaussian mixture models (Fig.~\ref{fig:mix}) and simply plotting the data (Fig.~\ref{fig:kinematic_split}).
    \item Using a period--age relation, this implies that there is a sharp increase in star formation in the NSD around $(8\pm1)\,\mathrm{Gyr}$ which we have identified with the time of bar formation. This feature is present irrespective of whether the selection function is included, whether the longitudinal proper motions are ignored and whether the proper motion errors are weakly modified. There is evidence for some NSD members older than $8\,\mathrm{Gyr}$ although as the bar kinematics are dominant in this regime this is somewhat uncertain.
    \item More flexible models of the NSD with period demonstrate weak gradients in the NSD scalelength, scaleheight and radial velocity dispersion with period consistent with the NSD forming inside-out and having undergone dynamical heating over time but the formation gradients have likely been washed out by significant dynamical mixing.
\end{enumerate}
To close, we will now discuss the significance of these results and highlight some limitations of the work and possible future extensions.

\subsection{Modelling limitations}
There are several assumptions in our modelling that could be relaxed in future work or with a better understanding of the NSD and its stellar tracers. For example, we have assumed that the age distribution of the background bar-disc `contamination' model is independent of distance and that the velocity distributions at a given location are independent of age. It is well known that velocity dispersion-age-position correlations exist in the disc and the bar-bulge \citep[e.g.][]{Hasselquist2020, Sharma2021}. Too much freedom to the background model will remove our constraining power as the background can become confused with the NSD component, so more prior information from the literature and other datasets would be required, or a more detailed understanding of the distance distribution of our sample employed.

Our discussion of the selection function highlighted that different Mira variable populations could trace different $K_s$ period--luminosity relations \citep{Sanders2023}. The cause for this is unclear but the first-order effects might be related to metallicity \citep{Trabucchi2019}. Without detailed spectroscopy, even approximate metallicities for our sample are not possible, but it seems likely there is some metallicity variation in our sample and in particular, a variation in mean metallicity between the NSD and bar populations \citep{Schultheis2021}. This adds a layer of complexity to the selection function where we may be losing the intrinsically brighter more metal-poor objects that make up the bar population in favour of the fainter metal-rich objects that are preferentially part of the NSD population. We have demonstrated that our conclusions are independent of selection effects so this effect is not a significant concern but it highlights that there is potential in future work to combine the models with spectroscopic datasets \citep[][\citetalias{Sormani2022}]{Fritz2021, Schultheis2021} to yield more powerful constraints.

\subsection{Comparison with other NSD studies}
\begin{figure}
    \centering
    \includegraphics[width=\columnwidth]{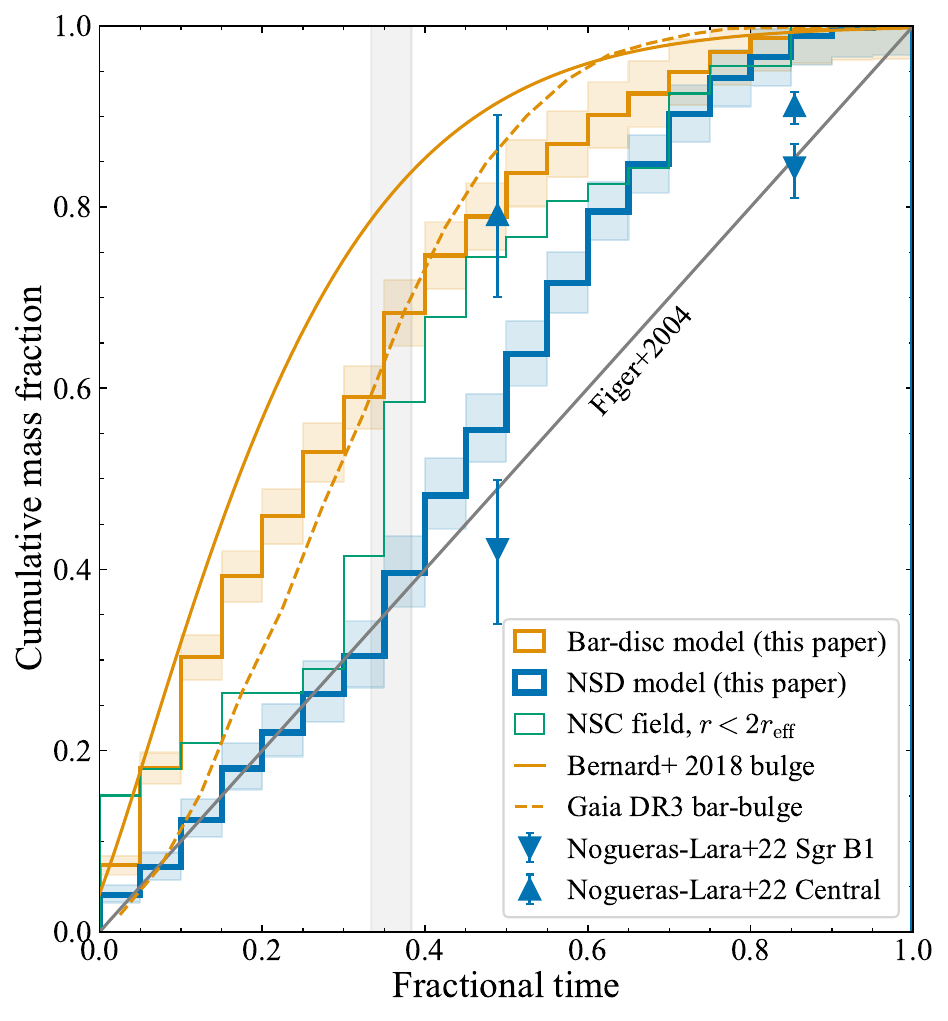}
    \caption{
    Comparison with other studies: the histograms show cumulative mass growth of the NSD (thick blue), bar-bulge (thinner orange) and the NSC region (thinnest green). The orange line shows the cumulative star formation history of the bar-bulge inferred by \protect\cite{Bernard2018}, the dashed orange line the cumulative bar-bulge star formation history inferred from Gaia DR3 Mira variables within $|\ell|<1.5\,\mathrm{deg}$ and $|b|<1.5\,\mathrm{deg}$, and the black errorbars are cumulative mass measurements for two regions of the NSD from \protect\cite{NoguerasLara2022}. The grey line is a uniform star formation history as advocated by \protect\cite{Figer2004}. The x-axis is fractional time (time/maximum time) to accommodate slightly different age scales (see text for details).}
    \label{fig:comparison}
\end{figure}
Our derived formation epoch for the NSD is consistent with the idea of a `middle-aged' to old Galactic bar as found in other studies \citep{Bovy2019, Nogueras-Lara2020, NoguerasLara2023, Wylie2022, Schoedel2023} and seems to strongly rule out any suggestion that the Milky Way bar is only a few Gyr old. We here discuss more critically the comparison of our result with other related work. Our derived star formation histories for the NSD and bar are compared to other studies in Fig.~\ref{fig:comparison}. We have opted to plot the results against fractional time to divide out small age systematics we will discuss below.

Our NSD star formation history is approximately consistent with the idea of an old burst as advocated by \cite{Nogueras-Lara2020, NoguerasLara2023} and \cite{Schoedel2023}. Somewhat different to those studies, we find weak evidence for very ancient ($\gtrsim10\,\mathrm{Gyr}$) populations in the NSD. Our derived star formation history is relatively continuous with age \citep{Figer2004}. Simulations of the nuclear stellar discs, such as those of \cite{Seo2019}, predict a highly bursty star formation history highly dependent on the large-scale supply of gas. However, when smoothed over the typical scatter in the Mira variable period--age relation or  $\sim1\,\mathrm{Gyr}$ \citep{Trabucchi2022, ZhangSanders2023}, these simulations appear relatively consistent with the results presented here. One feature of the \cite{Nogueras-Lara2020} picture we are missing is a recent $1\,\mathrm{Gyr}$ ago star formation burst in the NSD \citep[][finds this burst could constitute $\sim15$ per cent of the mass in the NSD]{Schoedel2023}. It is clear from the presence of very young stellar tracers that there is ongoing star formation in the NSD region \citep{Morris1996, Matsunaga2015, Henshaw2022} but the identification of these less recent significant star formation episodes comes primarily from the presence of secondary red clump stars \citep{Nogueras-Lara2020, Schoedel2023}.
Another indicator of recent star formation is C-rich Mira variables \citep{Boyer2013, Matsunaga2017, SandersMatsunaga2023}. The single star formation channel for these objects is through dredge-up making their production more effective for massive and/or metal-poor stars. However, binary formation channels are also likely \citep{Matsunaga2017, SandersMatsunaga2023}. \cite{Sanders2022} used colour--colour diagrams to demonstrate $\lesssim10$ per cent of the present sample is consistent with being C-rich but this number is somewhat uncertain due to extinction effects. It is uncertain whether this number of objects is consistent solely with production through binary channels or whether it indicates some recent star formation in the population. Another possibility for missing a significant recent star formation episode is that our Mira variable selection is lacking long-period objects. Long-period Mira variables are easily identified as they are high amplitude, but they could be missed due to significant dust embedding or very irregular light curves due to circumstellar dust. A final reason for the discrepancy at recent times could be from our simple conversion from Mira variable number density to star formation history and/or the employed Mira period--age relation is inappropriate for the youngest stars (our employed period--age relations are an extrapolation below $\sim 3\,\mathrm{Gyr}$). 

It is interesting to briefly consider the relationship between the NSD and NSC star formation histories. The NSC appears to predominantly contain stars that are older than $5\,\mathrm{Gyr}$ but on average younger than the bar-bulge \citep[see the summary and references given by][]{Neumayer2020}. The recent study of \cite{Chen2023} argued the NSC has an age of $5^3_{-2}\,\mathrm{Gyr}$, younger than previous estimates due to their more flexible modelling of the metallicity distribution. Interestingly, this is very similar to the mean age of the NSD from our analysis. In Fig.~\ref{fig:comparison} we show the cumulative mass fraction from the $51$ Mira variables we have within a projected distance of $2r_\mathrm{eff}=8.4\,\mathrm{pc}$ from the NSC centre where we notice an uptick at exactly the bar formation time proposed in this work. These two pieces of evidence suggest a close evolutionary link between the NSD and the NSC \citep{NoguerasLara2023b}.

Although not the focus of the work, our inferred bar star formation history is approximately consistent with that presented by \cite{Bernard2018} using HST data in several bar-bulge fields. Our age scale saturates around $12\,\mathrm{Gyr}$ whilst the age scale of \cite{Bernard2018} extends to $14\,\mathrm{Gyr}$. This is suggestive of a $12/14\times8\,\mathrm{Gyr}\approx1\,\mathrm{Gyr}$ systematic in our bar age estimate. For this reason, we have opted to plot the fractional time in Fig.~\ref{fig:comparison}. Even accounting for this, our bar star formation history is biased slightly young with respect to the results of \cite{Bernard2018} which could be interpreted as the results of fractionation when the bar formed \citep{Debattista2017}. However, the models using only the Galactic latitude proper motion (see Fig.~\ref{fig:star_formation_history_comparison}) are biased older and more consistent with the results of \cite{Bernard2018} so this discrepancy appears within the systematics of our methodology. Finally, the central bar-bulge Mira variables in Gaia DR3 \citep{Lebzelter2022}, as identified with a $G$-band semi-amplitude cut of $0.45$, periods between $100$ and $1000$ days and in the region $|\ell|<1.5\,\mathrm{deg}$ and $|b|<1.5\,\mathrm{deg}$, produce a star formation history consistent with our bar-bulge star formation history with a slightly younger bias possibly due to foreground disc contamination.

\subsection{Alternative NSD formation scenarios}
We have weak evidence that the NSD is older than the $\sim8\,\mathrm{Gyr}$ star formation burst signal as some fraction of the oldest stars are attributed to the NSD. This could reflect contamination in the sample or shortcomings of the bar-disc model as discussed above, or could genuinely reflect a much more ancient NSD that had a `slow start'. A slow-start picture is somewhat opposed to the expectation from simulations \citep{Baba2019} where a significant burst in star formation is expected at the time of formation. However, potentially the Galaxy was relatively gas-poor at these early times \citep[although this seems unlikely,][]{Daddi2010} and only at some later epoch (possibly coinciding with a merger event) was significantly more gas accreted leading to a star formation burst. This discussion highlights that our bar age estimate is only ever going to be a lower bound as gas is required in the Galaxy at the time of bar formation and then some time is required for its transport to the centre. Another interpretation is that an ancient NSD formed after an early bar formation, the bar was destroyed (again possibly related to merger events) and then later a second bar formed. This idea is somewhat supported by the differing NSD kinematics for stars younger and older than $8\,\mathrm{Gyr}$. However, there is limited evidence for the destruction of bars after formation both theoretically and observationally \citep[see discussion in Section V.B.6 of][although see recent simulation work from \cite{Bi2022} and \cite{Cavanagh2022}]{SellwoodReview}. Therefore, we are well-justified in discussing a single bar formation epoch. Finally, it is possible that the old NSD material formed through a different channel. For instance, it could be produced by accretion through mergers that sink and remain intact deep into the Milky Way potential. From our calculations, it makes up a significant fraction of the mass ($30$ per cent so around $3\times10^8M_\odot$) which would require the order of hundreds of accreted clusters so seems unlikely. Another possibility is we are seeing the remnant of the very earliest `disc' that formed in the primordial Milky Way, an analogue of those seen at high redshift \citep{Kikuchihara2020, Ono2023}.

\subsection{Future observational prospects}
One limitation of using VVV data to probe the Mira variables in the NSD region is that, due to saturation, for $K_s\lesssim10.5\,\mathrm{mag}$ we only detect those NSD Mira variables with significant circumstellar and/or interstellar extinction. Additionally, this means our sample is strongly confined to high-extinction regions in the Galactic plane. In the presented modelling, we have included this effect and demonstrated it has a relatively weak impact on the conclusions. Ideally, we would use \emph{all} Mira variables in the NSD. However, this requires both their identification from multi-epoch data and measured proper motions. One option is Gaia: 
unfortunately due to the effects of extinction the Gaia DR3 Mira variable sample \citep{Lebzelter2022}
is confined to $|b|\gtrsim0.5\,\mathrm{deg}$ where the NSD stars are expected to only contribute $\lesssim10$ per cent relative to the background bar-bulge (see Fig.~\ref{fig:data}). Therefore, there is a `gap' in both the Mira variable on-sky and magnitude ($9\lesssim K_s\lesssim10.5$) distributions between the Gaia stars and the VIRAC sample presented here. This is somewhat filled in by the \cite{Matsunaga2009} sample, but these lack proper motion information. One promising future candidate for filling this gap is JASMINE \citep{KawataJASMINE2023}. The Galactic Centre Survey is designed to cover exactly this on-sky and magnitude range ($9.5\lesssim H_\mathrm{w}\lesssim14.5$) so if there are Mira variables there, JASMINE should find them. Finally, it is possible that the Roman telescope could reach down to the main sequence turn-off in the NSD \citep{Terry2023} although confusion, crowding and foreground/background subtraction may still be challenging. This would provide an independent way to date the different components of the NSD and also give a way to measure the age-metallicity variations in the NSD, an option not possible with Mira variables and currently only possible through combination with spectroscopic datasets \citep{Fritz2021, Schultheis2021, Sormani2022}. Our results demonstrate the great promise of further pursuing more sophisticated modelling of the limited NSD datasets as well as future surveys and projects to better understand this important but hard-to-study region of the Galaxy.

\subsection{The timeline of the Galaxy and the wider context} 
We close by returning to our initial goal: placing the bar formation epoch in the timeline of the Milky Way. We have presented evidence that the Milky Way bar formed $\sim8\,\mathrm{Gyr}$ ago. Within the expected systematic uncertainties of both estimates ($\sim1\,\mathrm{Gyr}$), this could place the bar formation time close to the infall of the Gaia-Sausage-Enceladus merger around $9-10\,\mathrm{Gyr}$ ago \citep{Belokurov2018, Belokurov2020, Bonaca2020} such that potentially the Milky Way has a tidally-induced bar \citep{Lokas2014}. However, it is also likely that the early disc growth \citep{BelokurovKravtsov2022, Semenov2023, Dillamore2023, Khoperskov2023} and relative dynamical quietness after the early mergers in the Milky Way gave rise to conditions in the disc that were more conducive to bar formation. Either hunting for observational signatures that the Milky Way bar is tidally induced \citep{Miwa1998} or more accurate timing of both the bar formation and the merger time of GS/E will be needed to answer this question more concretely.
It is also intriguing that around this $\sim8-10\,\mathrm{Gyr}$ age the Milky Way transitions from thick disc dominance to thin disc dominance. There is then the suggestion that the bar formation plays a role in this population separation \citep{Khoperskov2018} or possibly as with the merger discussion both are driven by the same root cause \citep{Grand2020}. Finally, measurement of the bar age can be combined with the observations of a presently slowing bar \citep{Chiba2021a, Chiba2021b} to approximately extrapolate to find the pattern speed of the bar at formation. This is ambitious and would require detailed modelling through simulations but may reveal insights into the relative importance of dark matter and gas for angular momentum transfer over the lifetime of the Galaxy. In conclusion, the synergy between high-redshift studies, large cosmological simulation suites and resolved stellar studies in the Milky Way of the type presented here are beginning to present a coherent picture of bar formation across the Universe.

{
\small
\section*{Acknowledgements}
JLS acknowledges the support of the Royal Society  (URF\textbackslash R1\textbackslash191555). 
DK acknowledges the support of the UK's Science \& Technology Facilities Council (STFC grant ST/S000216/1, ST/W001136/1) and MWGaiaDN, a Horizon Europe Marie Sk\l{}odowska-Curie Actions Doctoral Network funded under grant agreement no. 101072454 and also funded by UK Research and Innovation (EP/X031756/1). 
MCS acknowledges the financial support of the Royal Society (URF\textbackslash R1\textbackslash 221118). 
DM gratefully acknowledges support from the ANID BASAL projects ACE210002 and FB210003, from Fondecyt Project No. 1220724, and from CNPq Brasil Project 350104/2022-0
We thank the organisers and contributors of the Galactic Bars 2023 conference for their excellent updates on the interesting field of bar formation. 
This work has made use of data from the European Space Agency (ESA) mission
{\it Gaia} (\url{https://www.cosmos.esa.int/gaia}), processed by the {\it Gaia}
Data Processing and Analysis Consortium (DPAC,
\url{https://www.cosmos.esa.int/web/gaia/dpac/consortium}). Funding for the DPAC
has been provided by national institutions, in particular the institutions
participating in the {\it Gaia} Multilateral Agreement. Based on data products from observations made with ESO Telescopes at the La Silla or Paranal Observatories under ESO programme ID 179.B-2002. This paper made use of
\textsc{numpy} \citep{numpy},
\textsc{Jax} \citep{jax},
\textsc{Numpyro} \citep{numpyro1,numpyro2},
\textsc{scipy} \citep{scipy},
\textsc{matplotlib} \citep{matplotlib},
\textsc{seaborn} \citep{seaborn}, \textsc{pandas} \citep{pandas1},
\textsc{astropy} \citep{astropy:2013,astropy:2018} and \textsc{Agama} \citep{Vasiliev2019}.

\section*{Data Availability}
The Mira variable dataset from the work of \cite{Sanders2022} will be made available along with the proper motions via Vizier. Code related to this project is available at \url{https://github.com/jls713/mira_nsd}.
}



\bibliographystyle{mnras}
\bibliography{bibliography.bib} 



\appendix
\section{Selection function of the Mira variable sample}\label{appendix:selection_function}
In equation~\eqref{eqn::highlevel}, $S(\ell,b,s, P)$ is the selection function, which is the fraction of stars at each $(\ell,b,s, P)$ entering our sample. We compute this quantity as
\begin{equation}
\begin{split}
    S(\ell,b,s,P) = S(K_s(\ell,b,s,P)&) \\=S\Big(M_{K_s,\mathrm{NSD}}(P)&+5\log_{10}(100 [s/\mathrm{kpc}])\\&+A_{K_s}(\ell,b,s)\Big).
\end{split}
\label{eqn::selection_function}
\end{equation}
This expression consists of several components. $S(K_s)$ is the completeness curve given by \cite{Sanders2022}, which is largely governed by the failure to detect bright objects due to saturation in VVV. \cite{Sanders2022} assessed the impact of this effect using both artificial star tests and a comparison with the sample of \cite{Matsunaga2009}. The two methods produce similar results that are well approximated by the simple functional form
\begin{equation}
    S(K_s) = 0.45(1+\tanh[2.5(K_s-10.3)]),
\label{eqn::sk_eqn}
\end{equation}
i.e. the sample is maximally around $90$ per cent complete and stars with $K_s\lesssim 10.3$ have a significant probability to be missed. There are an insufficient number of Mira variables from \cite{Matsunaga2009} to fully assess the completeness at the faint end. The search of \cite{Sanders2022} did not impose any explicit $K_s$ cuts at faint magnitudes but period--Wesenheit magnitude cuts were employed by \cite{Sanders2022} to clean the sample (see their figure D1). These cuts are approximated by removing model star particles beyond $\sim 16\,\mathrm{kpc}$ (see Section~\ref{subsection:contaminant_model}). The inclusion of a faint limit taper in $S(K_s)$ at $K_s\approx13$ produces minimal changes to the selection-function-weighted bar model.

$A_{K_s}(\ell,b,s)$ is the 3D extinction map from \cite{Schultheis2014}. We use a coefficient $A_{K_s}/E(H-K_s)=1.306$ \citep{Sanders2022_ext} to convert the $E(H-K_s)$ values reported by \cite{Schultheis2014}. The resolution of the \cite{Schultheis2014} maps is $0.1\,\mathrm{deg}\times0.1\,\mathrm{deg}$ so highly variable extinction on small on-sky scales will not be well captured. However, we later will describe how we also consider the reported extinction scatter in our procedure, which mitigates this effect. Furthermore, we have found the results in this paper change very little when we instead use the 2d extinction maps of \cite{Sanders2022_ext} combined with the distance dependence of the \cite{Schultheis2014} map assuming the 2d map represents the extinction at the Galactic Centre distance.

\subsection{NSD Mira variable period--luminosity relation}
\begin{figure*}
    \centering
    \includegraphics[width=\textwidth]{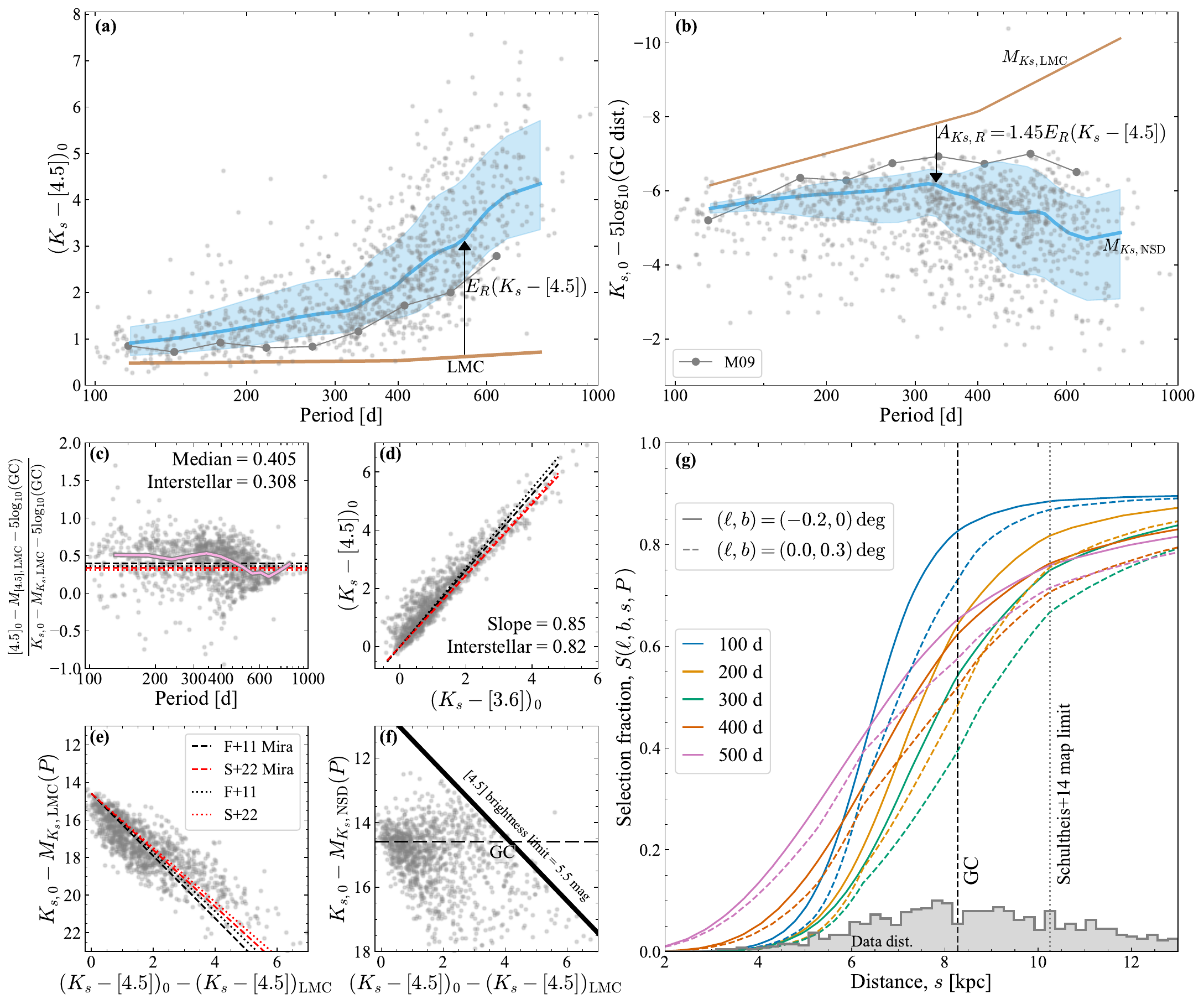}
    \caption{
    Illustration of the selection function calculation procedure. The grey points in all panels are the NSD Mira variable sample. \textbf{Panel (a)} shows the $(K_s-[4.5])$ colour corrected for interstellar extinction from the maps of \protect\cite{Schultheis2014} vs. period of the NSD Mira variable sample. The orange line is the measured relation for the LMC, and the cyan line and bracket show the median and $\pm1\sigma$ for the NSD Mira variable sample. The grey line with larger grey points shows the median for \protect\cite[][M09]{Matsunaga2009} Mira variables with $K_s<10.5$.
    \textbf{Panel (b)} shows similar but for dereddened absolute $K_s$ (computed assuming all stars are at the Galactic Centre distance). The cyan line and bracket have been computed by adjusting the LMC period--luminosity relation (orange line) by the residual dust extinction as measured by the colour excess in panel (a) and removing the extinction scatter and multi-epoch scatter for the bracket.
    \textbf{Panel (c)} shows the ratio of the residual reddening in $[4.5]$ and $K_s$ assuming all stars are at the Galactic centre distance. The pink line is the median and the four horizontal lines are different interstellar extinction calculations as labelled in panel (e). 
    \textbf{Panel (d)} shows $(K_s-[4.5])$ vs. $(K_s-[3.6])$ dereddened by the \protect\cite{Schultheis2014} maps with the interstellar extinction laws overplotted.
    \textbf{Panel (e)} shows $K_s$ dereddened by the \protect\cite{Schultheis2014} maps minus the absolute magnitude relation for the LMC, $M_{K_s,\mathrm{LMC}}$ (as displayed in panel (b)), against the colour excess from panel (a).
    \textbf{Panel (f)} shows the distance modulus computed from $K_s$ dereddened by the \protect\cite{Schultheis2014} maps and the NSD period--luminosity relation, $M_{K_s,\mathrm{NSD}}$, from panel (b) against the colour excess from panel (a).
    Finally, \textbf{panel (g)} shows the selection fraction, $S(\ell,b,s,P)$, with distance for two representative on-sky positions (distinguished by the line styles) and five periods (distinguished by the colours). Shorter-period stars are fainter so less affected by saturation.
    The Galactic Centre distance is marked along with the maximum distance resolved by the extinction maps of \protect\cite{Schultheis2014}. The distance distribution of the NSD Mira variable sample is shown assuming $M_{K_s,\mathrm{NSD}}$ from panel (b).}
    \label{fig:sf}
\end{figure*}

$M_{K_s,\mathrm{NSD}}(P)$ is the period-luminosity relation of the Mira variables in our sample. As discussed extensively by \cite{Sanders2022}, after correcting the $K_s$ magnitudes of the Mira variable sample for interstellar extinction (in that case using the extinction maps from \citealt{Sanders2022_ext}), the Mira variables in the NSD region are significantly fainter than those in the solar neighbourhood and LMC, as assessed by the period--luminosity relation $M_{K_s,\mathrm{LMC}}(P)$ \citep[][see also \citealt{Glass2001} and \citealt{Lewis2023}]{Sanders2023} assuming the bulk of the observed stars are around the Galactic Centre distance. This can be seen in panel (b) of Fig.~\ref{fig:sf} and is slightly at odds with observations that suggest a near universality of the O-rich Mira variable $K_s$ period--luminosity relation based on the comparison of observations of the solar neighbourhood and the LMC \citep{Whitelock2008, Sanders2023} although it is possible metallicity effects for the super-solar NSD stars also play a role here \citep{Schultheis2021}. There are several other possible reasons for this discrepancy: (i) there is very significant background disc contamination -- this would require some Mira variables being at distances far beyond the expected edge of the far side of the Galactic disc \citep[e.g. see][]{Nikzat2022} and also the colours of the dereddened Mira variables do not agree with the expectation from the LMC (see panel (a) of Fig.~\ref{fig:sf}). Furthermore, \cite{Sanders2022} showed that several of the period--Wesenheit distributions for the NSD Mira variables (particularly those using mid-infrared bands) agree with LMC relations shifted to the Galactic Centre distance. These observations make this first hypothesis unlikely; (ii) the interstellar extinction is poorly estimated -- this might be possible due to the highly fractal nature and extinction maps being averaged over large regions biased towards low extinction objects whilst we are biased towards high extinction objects). However, panel (a) shows that the colour excess discrepancy with respect to the LMC relation is a function of period which suggests that it is not a result of poorly estimated interstellar extinction unless there is a significant bias of finding longer period Mira variables in more heavily extincted regions which is a possibility; (iii) there is significantly more circumstellar extinction around these Mira variables than observed for the solar neighbourhood and LMC samples. This is likely to be a significant effect for the longer-period Mira variables but for $P\lesssim300\,\mathrm{d}$ circumstellar extinction is anticipated to be low \citep{Ita2011}. It appears likely that the cause of the discrepancy is a combination of options (ii) and (iii): interstellar and circumstellar extinction effects. Fortunately, the approach we take is agnostic to how these two effects are separated. Furthermore, the main results in the paper are insensitive to the inclusion or exclusion of the selection function.

We can relate the period--luminosity relation of the Mira variables in our sample to those in the LMC as
\begin{equation}
M_{K_s,\mathrm{NSD}}(P) = M_{K_s,\mathrm{LMC}}(P) + A_{K_s,R}(P),
\end{equation}
where $A_{K_s, R}(P)$ is a residual level of interstellar plus circumstellar extinction after the interstellar extinction correction from \cite{Schultheis2014}. The residual extinction can be estimated by comparison of the colours of our sample to the period--colour relations for LMC Mira variables
\begin{equation}
\begin{split}
(K_s-c) &\equiv (K_s-c)_0 + E_I(K_s-c),\\
&=(K_s-c)_\mathrm{LMC}(P) + E_I(K_s-c) + E_R(K_s-c) ,
\end{split}
\end{equation}
where $c$ is some arbitrary magnitude, $(K_s-c)_\mathrm{LMC}(P)$ is the period--colour relation for the LMC, $E_R(K_s-c)$ is the residual reddening and $(K_s-c)_0$ is the dereddened colour accounting solely for the interstellar reddening, $E_I(K_s-c)$, using the \cite{Schultheis2014} maps evaluated at $s=8.275\,\mathrm{kpc}$ as described above and the extinction law from \cite{Sanders2022_ext}. We then estimate the residual extinction term as a function of period by computing a running median of the relation
\begin{equation}
E_R(K_s-c)(P) = (K_s-c)_0 - (K_s-c)_\mathrm{LMC}(P).
\end{equation}
We convert the residual reddening, $E_R(K_s-c)(P)$, into a residual extinction $A_{K_s,R}(P)$ using a choice of extinction coefficient, $C_c\equiv A_{K_s,R}/E_R(K_s-c)$, which we will discuss shortly. At each period $P$, we also naturally expect some variation in the circumstellar extinction properties (possibly through some metallicity variation). In a similar way to the computation of the period--luminosity relation, we compute its spread using the spread in the residual extinction via running medians in period as
\begin{equation}
\begin{split}
\sigma_{K_s,\mathrm{NSD}}(P)^2 = C_R^2\Big[\mathrm{Var}(K_s&-c) - \sigma^2_{E_I}(K_s-c)\\&- \Delta E_I^2(K_s-c) - \Delta^2(K_s-c)\Big],
\end{split}
\end{equation}
where $\mathrm{Var}(K_s-c)$ is the variance of data and $\sigma_{E_I}(K_s-c)$ is the extinction uncertainty reported by \cite{Schultheis2014} evaluated at a distance of $8.275\,\mathrm{kpc}$ and $\Delta E_I(K_s-c)$ is half the difference between the extinction at $s=6.275\,\mathrm{kpc}$ and $s=10.275\,\mathrm{kpc}$ (approximately the expected distance range of the sample). The extinction spread term is subdominant so this assumption is not very important for the overall selection function calculation. $\Delta(K_s-c)$ is the spread of the colour arising from the variability of the stars. We use a mean $K_s$ measurement so this is only significant for the magnitude $c$. The uncertainties in the measurements of $K_s$ and $c$ are small so we do not consider them as contributing to the scatter.  We put a $0.13\,\mathrm{mag}$ floor on $\sigma_{K_s,\mathrm{NSD}}$ as this is the scatter in the period--luminosity relation for O-rich LMC stars using the LMCNISS data \citep{Macri2015, Yuan2018, Sanders2023} although this choice of floor is not too important as it is much narrower than the width of $S(K_s)$ in equation~\eqref{eqn::sk_eqn}.

\subsection{Application of the selection function procedure}
This procedure is illustrated in panels (a) and (b) of Fig.~\ref{fig:sf} using the selection of stars described at the start of Section~\ref{sec:data} (not restricted to low latitude only). We choose $c=[4.5]$ as our reference magnitude. because it is available from Spitzer/GLIMPSE catalogues \citep{Ramirez2008, GLIMPSE} for the majority of our sample, and because, as $[4.5]$ is a redder band, the variability is small. Additionally, \cite{Sanders2022} demonstrated the period--Wesenheit magnitude relations using $K_s$ and $[4.5]$ for the sample were similar to the LMC and solar neighbourhood samples (particularly for short-period stars). This may be because circumstellar extinction effects become minimal around $3\,\mu\mathrm{m}$ \citep{Ita2011, Lewis2023}. The GLIMPSE dataset is built by averaging over two epochs. Comparison with the model-average $W2$ WISE measurements from \cite{Sanders2022} suggests the scatter arising from variability, $\Delta(K_s-[4.5])$, is around $0.3\,\mathrm{mag}$ with a weakly increasing trend towards longer periods.

In our procedure, we must adopt an extinction law for the Mira variables to connect the residual extinction $A_{K_s, R}$ to the residual reddening $E_R(K_s-[4.5])$, $C_{[4.5]}\equiv A_{K_s, R}/E_R(K_s-[4.5])$. As discussed in the previous section, this residual extinction is likely a combination of interstellar and circumstellar extinction so this is not a simple exercise. There have been several investigations into the properties of circumstellar dust around Mira variables in both the Galactic bar-bulge and the LMC in the $J$, $H$ and $K_s$ bands \citep{Matsunaga2005, Ita2011, Yuan2017, Lewis2023} with most studies concluding the circumstellar extinction law in these bands differ from the interstellar extinction law (note that \citealt{Matsunaga2005} argue the interstellar and circumstellar extinction laws are similar although only when considering shallower interstellar extinction laws now disfavoured for the Galactic bar-bulge region).

In panel (c) of Fig.~\ref{fig:sf} we show the ratio of the discrepancy of $[4.5]_0$ dereddened using the \cite{Schultheis2014} maps with respect to the LMC relation shifted to the Galactic Centre distance modulus relative to the same for $K_s$. This ratio should give the residual $A_{[4.5]}/A_{K_s}$ reddening. We find a relatively flat trend with a median around the expectation from interstellar extinction laws \citep{Fritz2011, Sanders2022_ext}. There is a slight trend for higher coefficients for shorter-period stars, but this could be due to the relative importance of circumstellar and residual interstellar extinction with period or due to the distance distribution of the sample. Assuming different mean distances to the sample produces strong trends in this plane giving confidence that both the residual extinction coefficient is near the interstellar expectation and the sample is mostly located around the Galactic Centre distance.
This is further corroborated by panel (d) showing $(K_s-[4.5])_0$ vs. $(K_s-[3.6])_0$ dereddened by the \cite{Schultheis2014} maps. This distribution has a slope very similar to the interstellar extinction law (within a few per cent) and panel (e) which shows the dereddened residual with respect to $M_{K_s,\mathrm{LMC}}$ (not correcting for an assumed distance modulus) vs. the colour residual with respect to the LMC relation. The sample is consistent with being around a fixed distance and reddened according to an extinction law approximately similar to the interstellar extinction law. If we assume all of the residual extinction arises from circumstellar extinction, we confirm the result of \cite{Lewis2023} that the circumstellar extinction law has a smaller $A_H/A_{K_s}$ than the interstellar extinction law for this sample. However, it appears that around the $K_s$, $[3.6]$ and $[4.5]$ wavelengths the two extinction laws are quite similar. Using the interstellar extinction coefficient to relate $A_{K_s,R}$ to $E_R(K_s-[4.5])$ gives the blue line in panel (b). We see it neatly lies over the data distribution (which we assumed for the purpose of plotting is all located at the Galactic Centre distance). This is further evidenced in panel (f) showing the distance modulus assuming $M_{K_s,\mathrm{NSD}}$ vs. the colour residual. The distribution is centred around the Galactic Centre distance with a bias arising from the lack of $[4.5]$ measurements for stars brighter than $5.5\,\mathrm{mag}$ due to saturation. This bias does not affect our selection function as we are considering the selection in $K_s$.

It is advantageous that the residual extinction law is similar to the interstellar extinction law as it is then irrelevant how we portion out colour excess to each effect provided we do so consistently. The argument of the selection function is
\begin{equation}
\begin{split}
K_s&=M_{K_s,\mathrm{LMC}}(P)+5\log_{10}(100 [s/\mathrm{kpc}]) \\&+ C_{[4.5]}\Big[(K_s-[4.5])-(K_s-[4.5])_\mathrm{LMC}(P)\Big]\\&
+A_{K_s} - C_{[4.5]}E_I(K_s-[4.5]),
\end{split}
\end{equation}
where if $C_{[4.5]}=A_{K_s}/E_I(K_s-[4.5])$ the interstellar part cancels. This is also true in the handling of the scatter term although one could consider making different assumptions about the scatter in colour excess arising from interstellar extinction variation along a given line-of-sight which might weakly affect the final scatter calculation. Multiplying the residual extinction coefficient $C_{[4.5]}$ by $0.9$ to better match the distribution in panel (e) of Fig.~\ref{fig:sf} produces negligible changes to the selection function. A disadvantage of our procedure is that residual issues with interstellar extinction are not considered as a function of on-sky location, only period.

For our selection function approach to be valid, $M_{K_s,\mathrm{NSD}}(P)$ must reflect the period--magnitude relation of \emph{all} Mira variables in the NSD region, not just those we have observed. This is naturally quite difficult to assess. In panels (a) and (b) we show $(K_s-[4.5])_0$ and $K_{s,0}-\mu_\mathrm{GC}$ both dereddened by the \cite{Schultheis2014} maps for the \cite{Matsunaga2009} sample, which contains many stars brighter than our sample and itself is subject to selection effects. As shown by \cite{Sanders2022}, the \cite{Matsunaga2009} sample is bluer and brighter than the sample inspected here suggesting our sample is not completely representative and we should be cautious extrapolating the properties of our Mira variable sample and $M_{K_s,\mathrm{NSD}}(P)$ to lower extinction regions.

With the period--luminosity relation, $M_{K_s,\mathrm{NSD}}(P)$, and its scatter, $\sigma_{K_s,\mathrm{NSD}}(P)$, defined, we can return to equation~\eqref{eqn::selection_function} to evaluate the selection function. We convolve $S(K_s)$ with $\sqrt{\sigma^2_{K_s,\mathrm{NSD}}+\sigma^2_{A_{K_s}}}$ to account for the scatter in both the interstellar and residual extinction. The properties of the selection function are shown in Fig.~\ref{fig:sf}. For fixed period, the selection function biases the sample towards more distant stars, particularly those in the background disc. Due to the scatter in the extinction, the differential selection effects across the NSD region are minimal, although this is likely also a reflection of the shortcomings of the low-resolution 3D extinction maps we are limited to using. When comparing different stars of different periods, we see that stars with periods around $300\,\mathrm{day}$ are most strongly affected by selection effects (this is approximately the turnover in $M_{K_s,\mathrm{NSD}}$ in panel (b) of Fig.~\ref{fig:sf}) with shorter and longer period stars more readily detected in VVV.

The selection effects do not appear to significantly affect the conclusions of our work. However, there is certainly scope to further understand the interstellar and circumstellar extinction effects on this sample. This may be possible through a more detailed study of the Gaia sample located at higher latitudes and lower extinctions.

\section{Additional posterior predictive checks}\label{sec:additional_posterior}
\begin{figure*}
\centering
    \includegraphics[width=0.98\textwidth]{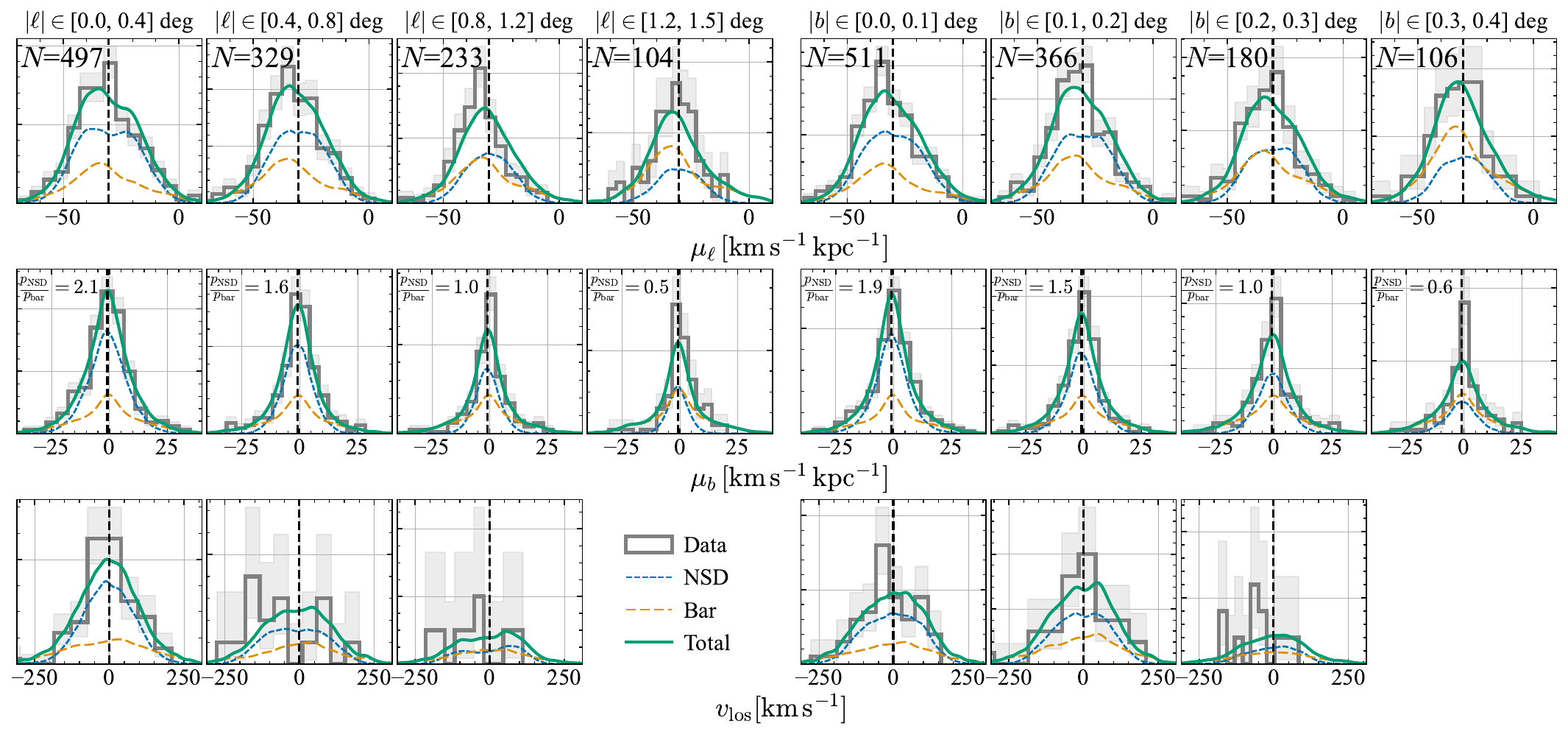}
    \caption{Velocity distributions for the data (grey with Poisson uncertainty shaded interval) compared to the best-fitting model split by Galactic longitude (left) and Galactic latitude (right). See the caption of Fig.~\ref{fig:veldists} for more information.}
    \label{fig:veldists_onsky}
\end{figure*}
In addition to posterior predictive checks split by period in the main body of the paper (see Fig.~\ref{fig:veldists}), we show in Fig.~\ref{fig:veldists_onsky} the mock samples split by on-sky location. This demonstrates the validity of the model fits to the data and demonstrates the decreasing dominance of the NSD with increasing $|\ell|$ and $|b|$. There is weak evidence of the model underpredicting the peak in $\mu_b$ which could arise in part from missing background disc stars in the model (possibly from slight oversmoothing in the kernel density estimates).

\section{Relation to bar age estimates for external galaxies}\label{sec:timer}
\begin{figure}
    \centering
    \includegraphics[width=\columnwidth]{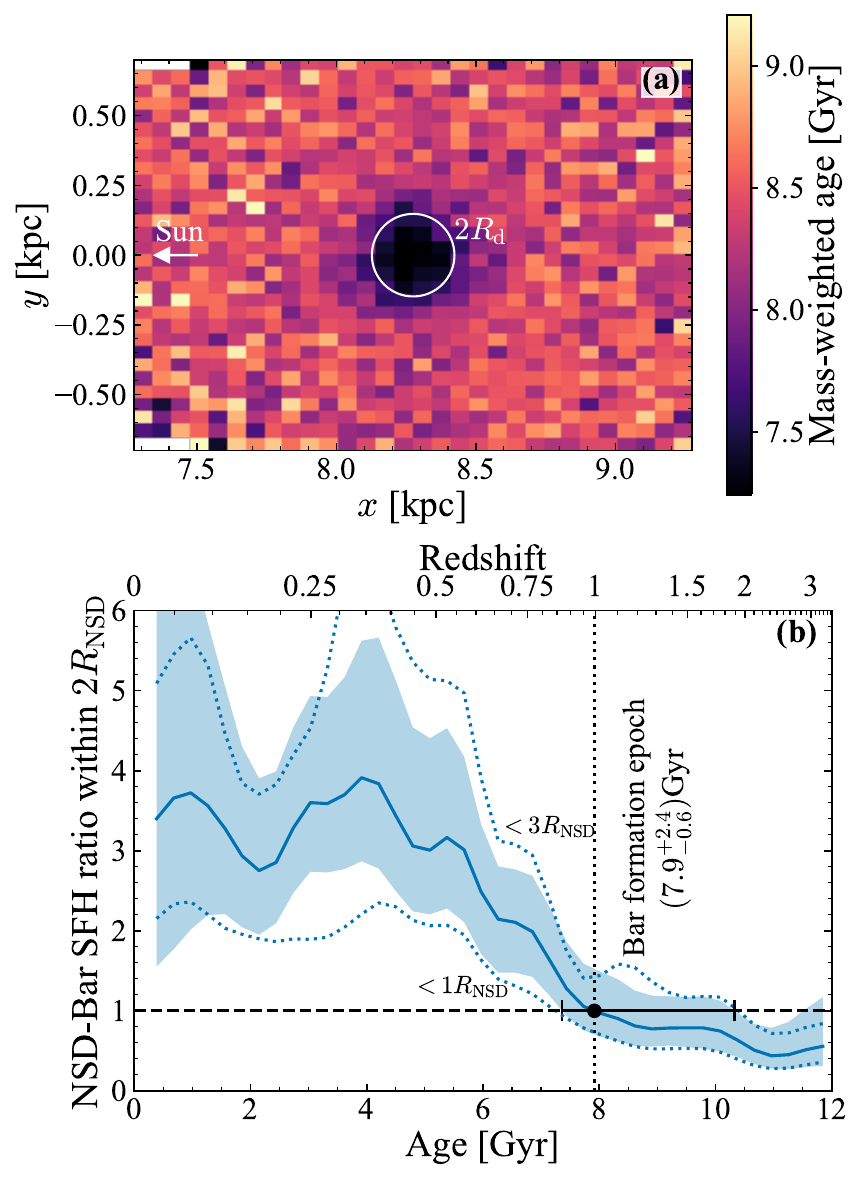}
    \caption{The Milky Way NSD as seen by an external observer: \textbf{Panel (a)} shows the mass-weighted age map of the inner regions of the Galaxy as seen by an observer located at infinity towards the Galactic North Pole. The white circle shows the region encompassing two scalelengths of the NSD and the Sun is located to the left outside the diagram. No extinction is considered. \textbf{Panel (b)} shows the ratio of the NSD to Bar star formation histories within a cylinder of $2R_\mathrm{NSD}$ (with uncertainty given by the shaded bracket and the two dotted lines the results within $1$ and $3R_\mathrm{NSD}$. Redshift is computed using the cosmological parameters from \protect\cite{Planck}. Following \protect\cite{deSaFreitas2023}, we define the bar formation epoch as the age at which the NSD is the dominant star-forming component within $2R_\mathrm{NSD}$ leading to the errorbar and annotated age for the Milky Way bar as inferred by an outside observer.}
    \label{fig:outside_view}
\end{figure}

We can connect our results to studies of measuring the bar formation epochs in other galaxies \citep{Gadotti2015,Gadotti2018,Gadotti2020,deSaFreitas2023,deSaFreitas2023b}. In Fig.~\ref{fig:outside_view} we display a mass-weighted age map of the inner few kpc as seen by an observer looking down on the disc of the Galaxy from infinity, the type of data one might obtain from an IFU. The NSD stands out as a slightly younger structure within the old bar. The difference in mean age is quite subtle ($\sim1\,\mathrm{Gyr}$) and only apparent within $2-3R_\mathrm{NSD}$. \cite{deSaFreitas2023,deSaFreitas2023b} present a methodology for constraining a bar's age from IFU data. As in our methodology, the bar contamination within the NSD region is significant, so \citeauthor{deSaFreitas2023} use the criterion for the bar formation epoch as the age where the NSD star formation history within a certain radius dominates over the bar i.e. $\int_{R<R_\mathrm{t}}\mathrm{d}A\,\mathrm{\Gamma}_\mathrm{NSD}/\int_{R<R_\mathrm{t}}\mathrm{d}A\,\mathrm{\Gamma}_\mathrm{bar}=1$. In the lower panel of Fig.~\ref{fig:outside_view}, we display this ratio for a radius of $R_\mathrm{t}=2R_\mathrm{NSD}=2\times74\,\mathrm{pc}$ (note here $R_\mathrm{NSD}$ is the `true' exponential scalelength of the disc and not the parameter of the DF) along with the scatter arising from uncertainty in $w(P)$ from our fiducial model. This method gives a bar age of $7.9^{+2.4}_{-0.6}\,\mathrm{Gyr}$ in line with our interpretation of bursts in the star formation in the NSD. Note that choosing a different radius $R_\mathrm{t}=R_\mathrm{NSD}$ or $R_\mathrm{t}=3R_\mathrm{NSD}$ produces a similar variance in the estimate as the uncertainty in $w(P)$ does. This age estimate agrees well with the estimates from the sharp rise in the NSD age distribution (see Table~\ref{tab:bar_age}) which, together with the tests on simulations from \cite{deSaFreitas2023}, gives good evidence of the self-consistency of this method of estimating the bar age.


\bsp	
\label{lastpage}
\end{document}